\documentclass[12pt]{article}
\usepackage{mathptmx} 
\usepackage{setspace} 
\usepackage{amsmath}
\usepackage{amsthm,amsfonts}

\usepackage{graphicx,psfrag,epsf}
\usepackage[plain,noend]{algorithm2e}
\usepackage{multirow}
\usepackage{xcolor}

\usepackage{enumerate}
\usepackage{natbib}
\usepackage{multibib}
\newcites{latex}{Additional References}
\usepackage{ccaption}
\usepackage{booktabs}
\usepackage{floatpag}
\usepackage{float}
\usepackage{url} 
\usepackage{makecell}
\usepackage{multicol}
\usepackage{multirow}
\usepackage{hyperref}
\usepackage{caption}
\DeclareCaptionFont{tenpt}{\fontsize{10pt}{10pt}\selectfont #1}
\newcommand{\blind}{1}

\numberwithin{equation}{section}

\newcommand{\R}{{\rm I}\kern-0.18em{\rm R}}
\newcommand{\N}{{\rm I}\kern-0.18em{\rm N}}
\newcommand{\Z}{{\rm Z}\kern-0.44em{\rm Z}}
\newcommand{\E}{{\rm I}\kern-0.18em{\rm E}}
\newcommand{\Prob}{{\rm I}\kern-0.18em{\rm P}}
\newcommand{\pr}{{\rm I}\kern-0.18em{\rm P}}
\newcommand{\1}{{\rm 1}\kern-0.24em{\rm I}}

\DeclareMathOperator*{\argmax}{arg\,max}
\usepackage{caption}

\begin{document}

\def\spacingset#1{\renewcommand{\baselinestretch}%
{#1}\small\normalsize} \spacingset{1}


\if1\blind
{
  \title{\large \bf Question-Score Identity Detection (Q-SID): A Statistical Algorithm to Detect Collusion Groups with Error Quantification from Exam Question Scores}
  \author{Guanao Yan,\hspace{.2cm} Jingyi Jessica Li
    \\
    Department of Statistics and Data Science,\\ University of California, Los Angeles\\\\
    Mark D. Biggin \\
    Biological Systems and Engineering Division,\\ Lawrence Berkeley National Laboratory, Berkeley\\}
  \date{}
  \maketitle
} \fi

\if0\blind
{
  \bigskip
  \bigskip
  \bigskip
  \begin{center}
    {\LARGE\bf Question-Score Identity Detection (Q-SID): A Statistical Algorithm to Detect Collusion Groups with Error Quantification from Exam Question Scores}
\end{center}
  \medskip
} \fi

\doublespacing
\vspace{-.4in}
\begin{abstract}
Collusion between students in online exams is a major problem that undermines the integrity of the exam results. Although there exist methods that use exam data to identify pairs of students who have likely copied each other’s answers, these methods are restricted to specific formats of multiple-choice exams. Here we present a statistical algorithm Q-SID that efficiently detects groups of students who likely have colluded, i.e., collusion groups, with error quantification. Q-SID uses graded numeric question scores only, so it works for many formats of multiple-choice and non-multiple-choice exams. Q-SID reports two false-positive rates (FPRs) for each collusion group: (1) \textit{empirical FPR}, whose null data are from $36$ strictly proctored exam datasets independent of the user-input exam data and (2) \textit{synthetic FPR}, whose null data are simulated from a copula-based probabilistic model, which is first fitted to the user-input exam data and then modified to have no collusion. On $34$ unproctored exam datasets, including two benchmark datasets with true positives and negatives verified by textural analysis, we demonstrate that Q-SID is a collusion detection algorithm with powerful and robust performance across exam formats, numbers of questions and students, and exam complexity. 
\end{abstract}
\singlespacing

\noindent%
\vspace{-.2in}
\vfill

\newpage
\spacingset{1.7} 
\section{Introduction}
\label{sec:intro}
Cheating on exams is a long-standing problem in academic testing (\cite{baird1980current, mccabe2005cheating, fox2008using}) and has increased with the rise in online testing because rigorous proctoring is difficult (\cite{watson2010cheating, fask2014online, diedenhofen2017pagefocus, dendir2020cheating, bilen2021online}). One of the most common means of cheating is collusion, in which small groups of students collaborate to copy each other's answers. Many methods have been proposed to detect collusion from exam data (\cite{angoff1974development, frary1977indices, belleza1989detection, jennings1996crime, wesolowsky2000detecting, mcmanus2005detecting, sotaridona2006detecting, andries2008detecting, disario2009applying, van2015bayesian, richmond2015detection, romero2015optimality, maynes2016detecting, fendler2018observing, lin2020catching}). These methods use the answers from multiple-choice exams---for example, choices \{a, b, c, d, or e\} or \{true or false\} for each question---and identify pairs of students who share an unusually high number of answers. However, these methods and studies have several limitations. 

First, these methods can only be used for multiple-choice exams, whereas many exams in STEM subjects include answers that vary from a word to several sentences, or include chemical formulae, diagrams, worked calculations, and lines of code. 

Second, it is difficult to estimate these methods' \textit{true positive rate} (\textit{TPR}, i.e., the faction of colluding students who are detected, also known as \textit{power}) because solid evidence of collusion is difficult to obtain from multiple-choice exam data, compared to exam data that include textual answers. In some cases, a subset of students identified by these methods was known to have cheated because they confessed (\cite{angoff1974development, lin2020catching}). In other cases, it was noted that students identified as likely to have colluded tend to be physically closer in a test center compared to students not considered likely (\cite{belleza1989detection, jennings1996crime, andries2008detecting, jennings1996crime, fendler2018observing, lin2020catching}). These results, however, likely underestimate the number of true positives. 


Third, the effectiveness of these methods has not been assessed for exams that have varying numbers of questions, numbers of examinees, or degrees of complexity. These methods’ general applicability is, thus, unclear. 

Motivated by these limitations of existing methods, we curated a large collection of exam datasets and developed a novel collusion detection algorithm, Question-Score Identity Detection method (Q-SID), to address the limitations as follows.

First, Q-SID is flexible regarding the input exam formats. Q-SID uses question scores (i.e., the graded scores for each question) to calculate a summary statistic, a per-student \textit{collusion score} (\textit{CS}), that identifies the students most likely to have colluded. Since questions in multiple-choice and non-multiple-choice exams all receive a numeric score, Q-SID is equally applicable to both exam types.  
    
Second, we curated ``ground-truth" datasets, including $36$ strictly proctored exams without collusion (null datasets with all true negatives) and two unproctored non-multiple-choice exams with collusion groups identified independently by manual textural analysis (datasets containing true positives). Note that true positives cannot be found by textural analysis in multiple-choice exam data (and that is why the existing methods cannot have TPR estimated accurately). Leveraging these ``ground-truth" data, we estimated Q-SID's TPR and \textit{false positive rate} (\textit{FPR}, i.e., the fraction of non-colluding students falsely detected). Moreover, to provide a more dataset-specific estimate of Q-SID's FPR, we generated synthetic null data from the exam dataset under investigation by sampling synthetic students as independent observations from a multivariate null distribution, whose variables are questions. To fit the multivariate null distribution on the exam dataset, we assumed every question to follow a multinomial distribution marginally, and we used the Gaussian copula to capture the pairwise correlations among questions. Such synthetic null data mimicked real data regarding questions' distributions but did not contain any collusion groups because students were sampled independently. Hence, for an input exam dataset, Q-SID provides two FPRs: (1) the \textit{empirical FPR} based on the $36$ real null exam datasets and (2) the \textit{synthetic FPR} based on the synthetic null data generated based on the input exam. 
    
Third, Q-SID is a general algorithm applicable to exams with a wide array of characteristics. We used subsampling or combining exams to vary the numbers of questions and students as well as the complexity of exams. Via comprehensive analyses, we set the default CS thresholds in Q-SID to maintain stable FPRs across exams of varying characteristics. 

The remainder of this article is structured as follows. Section 2 summarizes the input data format for Q-SID and the $70$ datasets (including $36$ strictly proctored exams and $34$ unproctored exams) used for developing Q-SID. 
Sections 3 and 4 cover the Q-SID methodology, including collusion group detection (Section 3) and error quantification (Section 4). Section 5 provides the power assessment of Q-SID. 
Section 6 demonstrates applications of Q-SID in real-world exam scenarios. Section 7 describes the Q-SID interactive web tool that facilitates its usage by instructors. Section 8 provides the discussion.

\section{The Data}\label{sec:data}
We used the data of $70$ exams taken at UC Berkeley or UCLA to develop Q-SID (Supplementary Table 1; Supplementary Data), including two benchmark datasets with true positives (students who colluded) and true negatives (students who did not collude) verified by textural analysis, $36$ proctored exams with all true negatives, and $32$ unproctored exams collected during the COVID-19 pandemic when exams had to be given online without effective proctoring. 

\subsection{Input data format} \label{subsec:input}
Q-SID requires the input exam data format to be a matrix, whose rows correspond to students with unique identifiers, columns correspond to questions, and entries correspond to question scores, each of which is a graded numeric score each student received for each question. 

\subsection{Benchmark datasets with true positives and true negatives} \label{subsec:data_benchmark}

A challenge faced by the existing multiple-choice-only collusion detection methods is that true positives cannot be reliably established using the exam data. In contrast, collusion on non-multiple-choice exams has long been judged by detailed manual analysis of written answers. Students who colluded have more similar answers than the remainder of the class, often because their answers share unusual errors (see Supplementary Material~\ref{subsec:supp_written} for more discussion). Hence, for non-multiple-choice exams, manual textural analysis, though time-consuming, can be used to define students who most likely colluded (i.e., true positives) and students who did not (i.e., true negatives).

We collected the two benchmark exam datasets from two unproctored, online exams in a UC Berkeley class during the pandemic, in which extensive collusion had occurred. Among the $263$ students who took both exams, we first identified pairs of students whose question scores were highly correlated, then from this set we defined $50$ true positives and $52$ true negatives via manual textural analysis, and we considered the remaining students as uncategorized. See Supplementary Material~\ref{subsec:supp_written} for our detailed process of defining true positives and negatives. 

Based on the two benchmark datasets, we constructed a summary statistic for each student, called the \textit{collusion score}, to inform the chance that the student colluded (Section \ref{subsec:metric}). We also used the true positives in the two benchmark datasets to decide two CS thresholds to find the students who colluded and the collusion groups (Section \ref{subsec:algorithm}).  

\subsection{Proctored exam datasets with all true negatives}\label{dat:true_neg}
We collected $36$ proctored exams taken between $2017$ and $2021$. $19$ of these exams are from a single STEM course taught three times per year at UC Berkeley, the same course for which the two benchmark exams were set. These $19$ proctored exams were each taken by over $300$ students and thus suitable for subsampling students to create exam datasets with varying sizes for robustness analysis of Q-SID. These $19$ proctored exams are referred to as the \textit{proctored control} and contain $6{,}847$ students in total. The remaining $17$ proctored exams were more heterogeneous, being taken by between $53$ and $575$ students in each exam, with a total of $3{,}969$ students. Together, the $36$ proctored exams contain a total of $10{,}816$ students, which we refer to as the \textit{proctored dataset}. For an input exam, Q-SID uses the proctored control to detect collusion groups, and it uses both the proctored control and proctored dataset to define the empirical FPR for each detected collision group (Section \ref{subsec:empFPR}).
    
\subsection{Unproctored exam datasets for Q-SID application} \label{subsec:data_add}
In addition to the two benchmark exams that were unproctored, there are $32$ unproctored online exams taken in $2020$ and $2021$, with the numbers of students between $54$ and $639$. In total, the $34$ unproctored exams have $10{,}526$ students. These unproctored exams will be used as application examples to demonstrate the efficacy of Q-SID. 

\section{Q-SID methodology: collusion group detection}\label{sec:method}
\subsection{Metrics for collusion detection}\label{subsec:metric}
We denote the \textit{question score matrix} by $\mathbf{X}\in \mathbb{R}_{\geq0}^{n\times p}$ with $n$ students as rows and $p$ questions as columns, where the \textit{question score} student $i$ received for question $s$, $X_{is} \in \mathbb{R}_{\geq0}$, is a non-negative real number. Using benchmark exam 2 as an example, there are $n=263$ students who answered $p=69$ questions. 
We define the \textit{test score} of student $i$ as the sum of student $i$'s question scores $\sum_{s=1}^p X_{is}$.

In the first step of Q-SID's collusion detection (Algorithm: Q-SID), every student is assigned a collusion score by the following procedure.

To begin, we define an \textit{identity score} (\textit{IS}) for each pair of students based on the students' question scores. The IS of students $i$ and $j \neq i$ is defined as the number of identical question scores shared by the two students:
\begin{eqnarray*}
\text{IS}_{ij}=\sum_{s=1}^p\mathbf{1}\left\{X_{is}=X_{js}\right\}.
\end{eqnarray*}

\begin{figure}[htbp]
\centering
\includegraphics[width=0.4\textwidth]{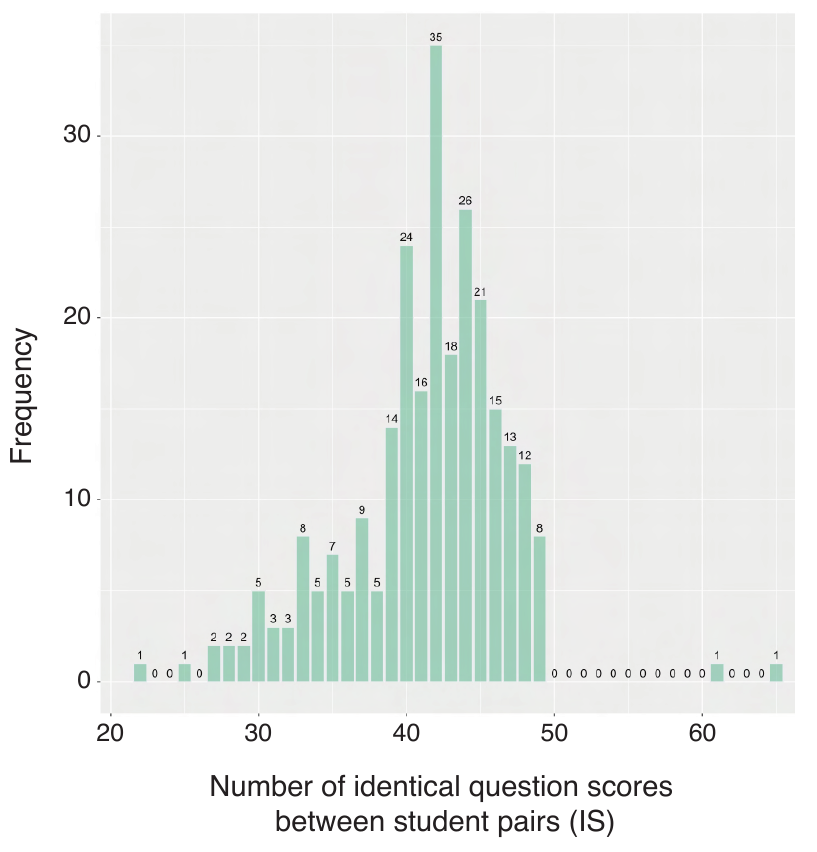}
\caption{Histogram of the ISs of one student with the other $262$ students who took benchmark exam 2. There are two large outliers of 61 and 65. \label{fig:IShist}}
\end{figure}

Hence, a student has $(n-1)$ ISs; for example, student $1$ has $\text{IS}_{12},\ldots,\text{IS}_{1n}$. Fig.~\ref{fig:IShist} shows the ISs between one student and each of the other $262$ students who took benchmark exam 2. There are two large outlier ISs of $61$ and $65$, which are notable since the exam only had $69$ questions.  Indeed, the three students corresponding to the two ISs were shown to have colluded by textural analysis of their exam answers, and all three confessed to cheating. 
	
Q-SID is based on the observation that large IS outliers are indicative of collusion. For this reason, each student $i$ is assigned a ``max IS" defined as $\underset{j\neq i}{\max}{\,}{\text{IS}_{ij}}$, the maximum of the  $(n-1)$ ISs student $i$ have with the remaining $(n-1)$ students. 
Accordingly, we define the student who has the max IS with student $i$ as student $i$'s \textit{1st partner}: ${i_1} = \underset{j\neq i}{\argmax} {\,} \text{IS}_{ij}$. Hypothetically, if a student has a large max IS, then the student might have colluded with his or her 1st partner. For the purpose of detecting collusion groups, we also define student $i$'s \textit{2nd partner} as $i_2 = \underset{j\neq i,i_1}{\argmax} {\,}\text{IS}_{ij}$. 

\begin{figure}[htbp]
\centering
\includegraphics[width=0.7\textwidth]{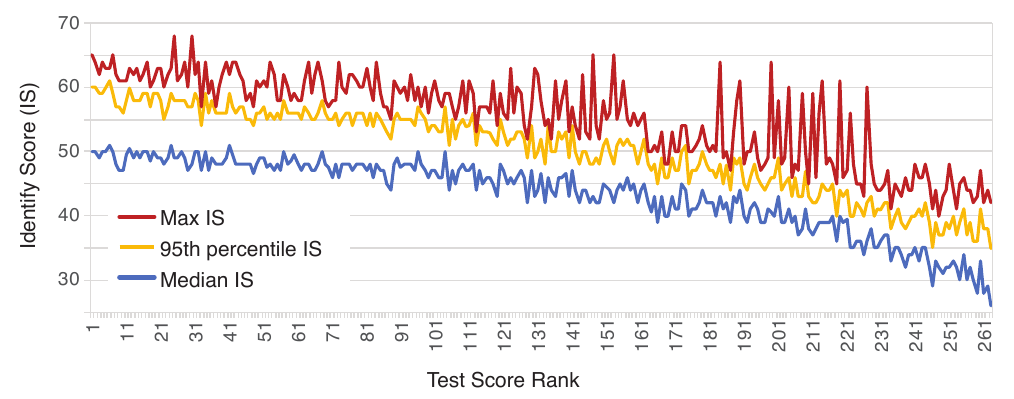}
\caption{Max IS, $95$th percentile IS, and median IS for each student who took benchmark exam 2, ranked by their test score. The student who received the highest test score was ranked number $1$. \label{fig:ISqtl}}
\end{figure}

However, this hypothesis is problematic because the max IS values suffer a systematic bias due to the dependence between a student's max IS and test score (i.e., sum of the student's question scores). The underlying reason for this dependence is that students who received higher test scores naturally had higher similarities in their question scores, leading to higher max IS values. Hence, a high max IS value alone cannot indicate collusion. Fig.~\ref{fig:ISqtl} (red line) shows that, in benchmark exam 2, the students who received high test scores generally have high max IS values.

To correct this bias, our idea is to find a baseline trend of IS values along test score ranks. The baseline should only reflect the bias that students who received top test scores tend to have high IS values, but not indicate collusion. If a student is among a group of students who received high test scores, then the student's $(n-1)$ IS values should contain some high values; hence, an upper quantile of the student's $(n-1)$ IS scores should be high. Motivated by this rationale, we examined two quantiles of each student's $(n-1)$ IS values: the median and the $95$th percentile. Based on benchmark exam 2, Fig.~\ref{fig:ISqtl} shows the trends of the two quantiles, which are highly correlated (with the Pearson correlation as $0.98$) and generally decrease as the students' test scores decrease, suggesting that the trends could be reasonable baselines. Since we deem it unlikely that students colluded in large groups, we do not think the median IS value is indicative of collusion. Hence, we use the median IS values to correct the bias of the max IS values.


Based on the rationale that the median IS values indicate the bias of the max IS values, we define student $i$'s \textit{identity metric} (\textit{IM}) as
\begin{eqnarray*}
\text{IM}_i=\max_{j\neq i}{\text{IS}_{ij}} - \underset{j\neq i}{\text{median}}{\,}{\text{IS}_{ij}}.
\end{eqnarray*}
Then we used the Pearson correlation to examine the bias removal effect. The correlation was $0.87$ between the median IS values and the max IS values, again confirming the existence and magnitude of biases in the max IS values. However, the correlation dropped to $0.06$ between the median IS values and the IM values, indicating that the biases were largely removed in the IM values.


\begin{figure}[htbp]
\centering
\includegraphics[width=0.75\textwidth]{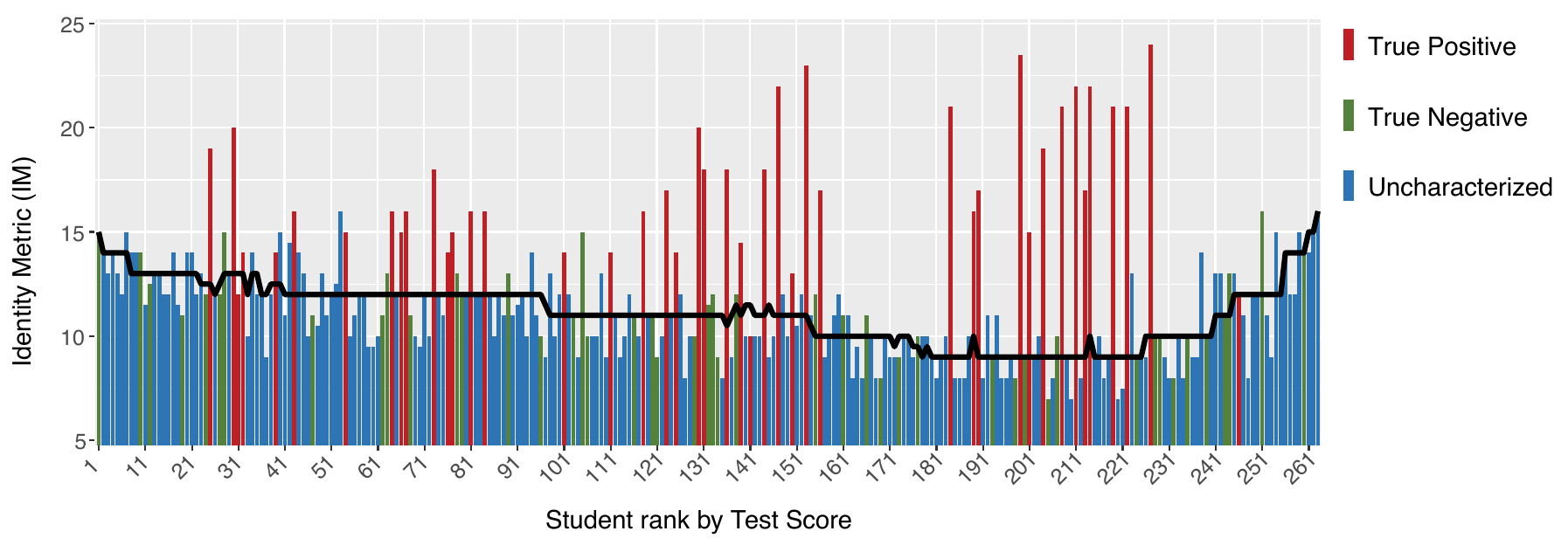}
\caption{Identity metrics (IMs) for benchmark exam 2, ranked by student test score. \label{fig:IM}}
\end{figure}

However, Fig.~\ref{fig:IM} suggests that the IM values need additional correction on benchmark exam 2. We observed that the local median IM values had a slight decline along the test score ranks from $1$ to $220$ but exhibited a rise for the bottom-ranked students. A possible reason is that the bottom-ranked students had many question scores as zeros, thus making their max IS values and median IS values behave differently from the other students. Hence, we further defined student $i$'s \textit{collusion score} (\textit{CS}) as
\begin{eqnarray*}
\text{CS}_i=\text{IM}_i/\left(\text{local\ median\ IM}_i\right),
\end{eqnarray*}
where the $\text{local\ median\ IM}_i$ is the median of the IMs of the immediate neighbors of student $i$ based on the test score rank (see Supplementary Material~\ref{subsec:supp_medIM} for the neighborhood size). 



\begin{figure}[htbp]
\centering
\includegraphics[width=0.75\textwidth]{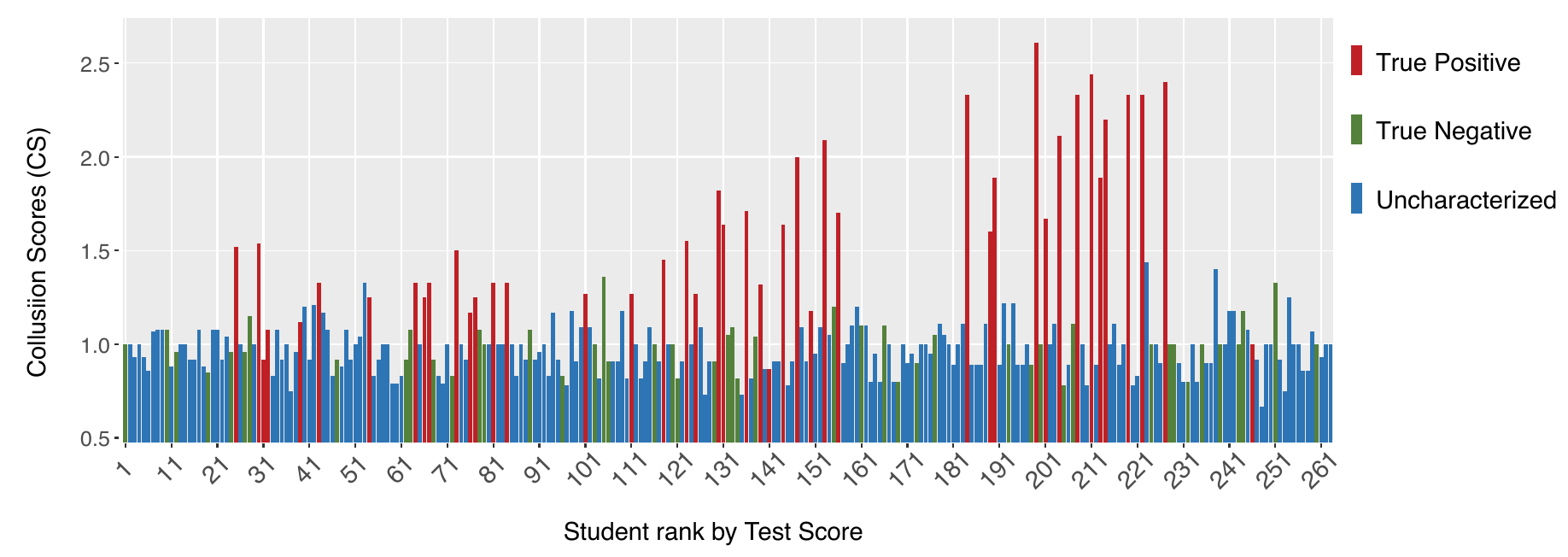}
\caption{Collusion scores (CSs) for benchmark exam 2, ranked by student test score. \label{fig:CS}}
\end{figure}

Fig.~\ref{fig:CS} shows the CS values of the students ordered by test score rank for benchmark exam 2. In both Figs.~\ref{fig:IM} and \ref{fig:CS}, red colored bars represent the students known to have colluded based on forensic analysis of written answers (i.e., true positives). In contrast, green colored bars highlight the students who have not colluded because no unusual similarities were spotted in their written answers (i.e., true negatives). In addition, blue colored bars show the students whose written answers have not been forensically analyzed (i.e., uncharacterized students). Inspecting the true negatives and uncharacterized students, we found no obvious bias in the CS values for the test score ranks, confirming that CS is a better metric than IM. As a result, the true positives became more distinct from the true negatives and uncharacterized students in terms of CS values (Fig.~\ref{fig:CS}) than IM values (Fig.~\ref{fig:IM}).


Note that before we calculated the CS for each student, we had a data preprocessing step to exclude (1) the students whose test scores were less than $5\%$ of the highest test score in the exam and (2) the repeated entries of the same student in the exam data (see Supplementary Material~\ref{subsec:supp_CSerrs} for detail). 

\subsection{Detection of collusion groups}\label{subsec:algorithm}
In steps 2 and 3 of Q-SID (Algorithm: Q-SID), given the intuition that the students with the highest CS values are those most likely to have colluded, we developed an algorithm to define collusion groups based on students' CS values.

First, step 2 of Q-SID identifies non-overlapping \textit{provisional collusion groups}, each of which is a set of students. Specifically, Q-SID first identifies all $\{$student, 1st partner$\}$ sets such that both students in a set have $\text{CS} \geq c_1$ and at least one student has $\text{CS} \geq c_2$, with the two thresholds $c_2 > c_1 > 0$.  The choice of threshold values for $c_1$ and $c_2$ will be discussed in Subsection~\ref{subsec:threshold}. Then, Q-SID combines the identified sets that overlap into $K$ non-overlapping provisional groups $\mathcal{T}_1,\ldots,\mathcal{T}_K$, which are ordered by the maximum CS of the students in each group, in a decreasing order.

Second, in step 3 of Q-SID,  for each of $k=1,\ldots,K$, if at least two members of a provisional group $\mathcal{T}_k$ share a 2nd partner that is a member of later provisional group $\mathcal{T}_s,~k<s\le K$, then $\mathcal{T}_s$ is merged into $\mathcal{T}_k$. After this merging, the remaining groups are designated as \textit{collusion groups} $\mathcal{G}_1, \ldots, \mathcal{G}_M$, with $M \le K$. 

Finally, for each collusion group $\mathcal{G}_m$, we define its maximum CS (maxCS) value $\mathrm{maxCS}_m$ as the maximum of the CS values of the students in it, $m=1,\ldots,M$. Due to the definition of collusion groups, all collusion groups have maxCS values at least $c_2$.



\subsection{Choice of CS value thresholds}\label{subsec:threshold}

Based on the two benchmark exams with true positives (i.e., students with confirmed collusion), we determined the CS value thresholds, $c_1$ and $c_2$, in an empirical, exam-size-dependent way based on (1) the two benchmark exams, for which $263$ students took both, and (2) the $19$ proctored control exams, which have $307$--$442$ students per exam, with a total of $6{,}847$ students we assumed to have not colluded. We calculated the CS value of each student within an exam and assumed that the CS values of the $6{,}847$ students in the proctored control exams represented an \textit{empirical null distribution} of CS values.

\begin{figure}[htbp]
\centering
\includegraphics[width=0.6\textwidth]{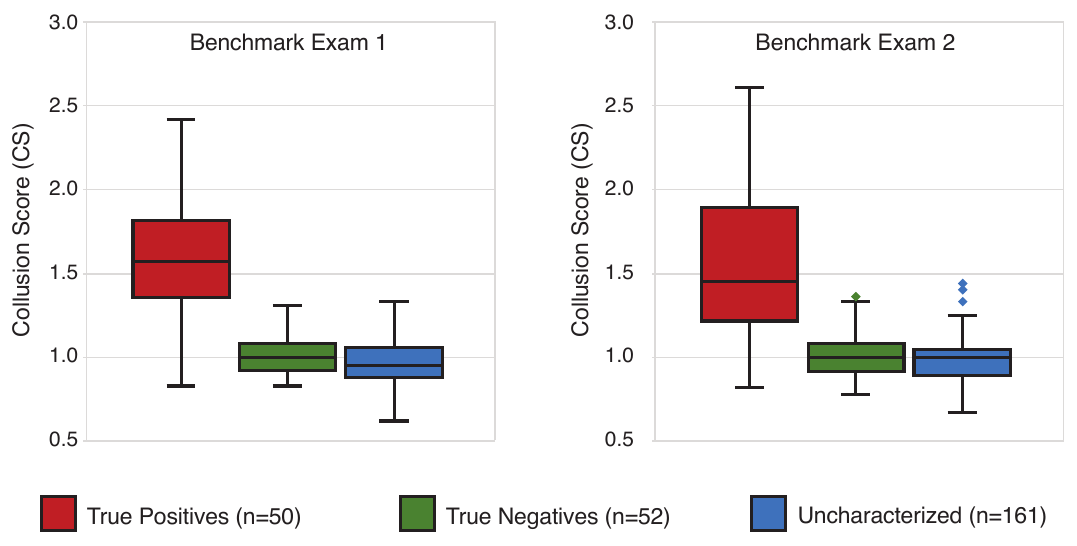}
\caption{The distributions of CS values for true positives, true negatives, and uncharacterized students who took the both benchmark exams. The number of students in each of the three sets is marked. \label{fig:CS_TP}}
\end{figure}

For an input exam with more than $250$ students, we used the two benchmark exams to determine the $c_1$ and $c_2$ thresholds as $c_1 = 1.23$ and $c_2 = 1.50$. Our choice of the thresholds was based on (1) the CS values of the true positives, true negatives, and uncharacterized students (Fig.~\ref{fig:CS_TP}) and (2) the CS values of the students' 1st partners (Supplementary Tables 2--3). Using these thresholds, Q-SID identified $46$ students to have likely colluded in at least one of the two benchmark exams. Among the $46$ identified students, $44$ students were confirmed as true positives (given the total of $50$ true positives, the true positive rate was $88.00\%$), and $2$ students were uncharacterized. 

Realizing that the CS value distribution depends on the exam size (i.e., number of students), we pre-calculated the thresholds $c_1$ and $c_2$ for an array of exam sizes: 15, 20, $\ldots$, 45, 50, 60, $\ldots$, 240, 250 by a subsampling-and-quantile-matching approach consisting of three steps. First, we converted the benchmark-exam-thresholds $c_1 = 1.23$ and $c_2 = 1.50$ to empirical null cumulative distribution function (CDF) values as $95.55\%$ and $99.71\%$ based on the empirical null distribution of CS values obtained from the $19$ proctored control exams (Table~\ref{tab:CS_thresh}; Fig.~\ref{fig:CS_thresh}). Second, for each exam size $n$ in $\{15, 20, \cdots, 45, 50, 60, \cdots, 240, 250\}$, we downsampled each of the $19$ proctored control exams to size $n$, calculated the CS value of each student within a downsampled exam, and repeat the downsampling and CS calculation for $100$ times. Then, we pooled the $(19 \times n \times 100)$ CS values to create an empirical null distribution for the exam size $n$. Third, we defined the $c_1$ and $c_2$ thresholds for the exam size $n$ as the $95.55\%$ and $99.71\%$ quantiles of the empirical null distribution.

Then, for an input exam with no more than $250$ students, we rounded the exam size to the closest integer $n$ in the set of $\{15, 20, \cdots, 45, 50, 60, \cdots, 240, 250\}$. Then we used the $c_1$ and $c_2$ thresholds for exam size $n$ for this input exam.




\section{Q-SID methodology: error quantification}\label{sec:FPR}
For any collusion detection method, a challenge is to obtain an accurate estimate of the FPR for each detected collusion group. If the estimate is poor or varies with the exam format, it would not be trusted by instructors, students, or academic administrators. 
To enhance the statistical rigor for collusion detection, Q-SID uses two types of null data (i.e., control exams) to evaluate the FPR of each detected collusion group: (1) empirical null data (i.e., proctored exams) and (2) synthetic null data (i.e., synthetic control exams). Below we introduce each type of null data and the corresponding definition of the FPR.

\subsection{Empirical FPR}\label{subsec:empFPR}
We defined the \textit{empirical FPR (empFPR)} for each collusion group based on the \textit{empirical null data} (also referred to as the \textit{proctored dataset}), consisting of $36$ proctored exams taken between the years $2017$ and $2021$ and containing $10{,}816$ students. Among the $36$ exams, 19 proctored control exams, consisting of $6{,}847$ students, were from the same course for which the two benchmark exams were set. We assumed the $10{,}816$ students in the proctored exams to have not colluded. Therefore, we defined the false positives as the students identified by Q-SID as likely not to have colluded in these exams (Section \ref{dat:true_neg}). 

\begin{figure}[htbp]
\centering
\includegraphics[width=0.8\textwidth]{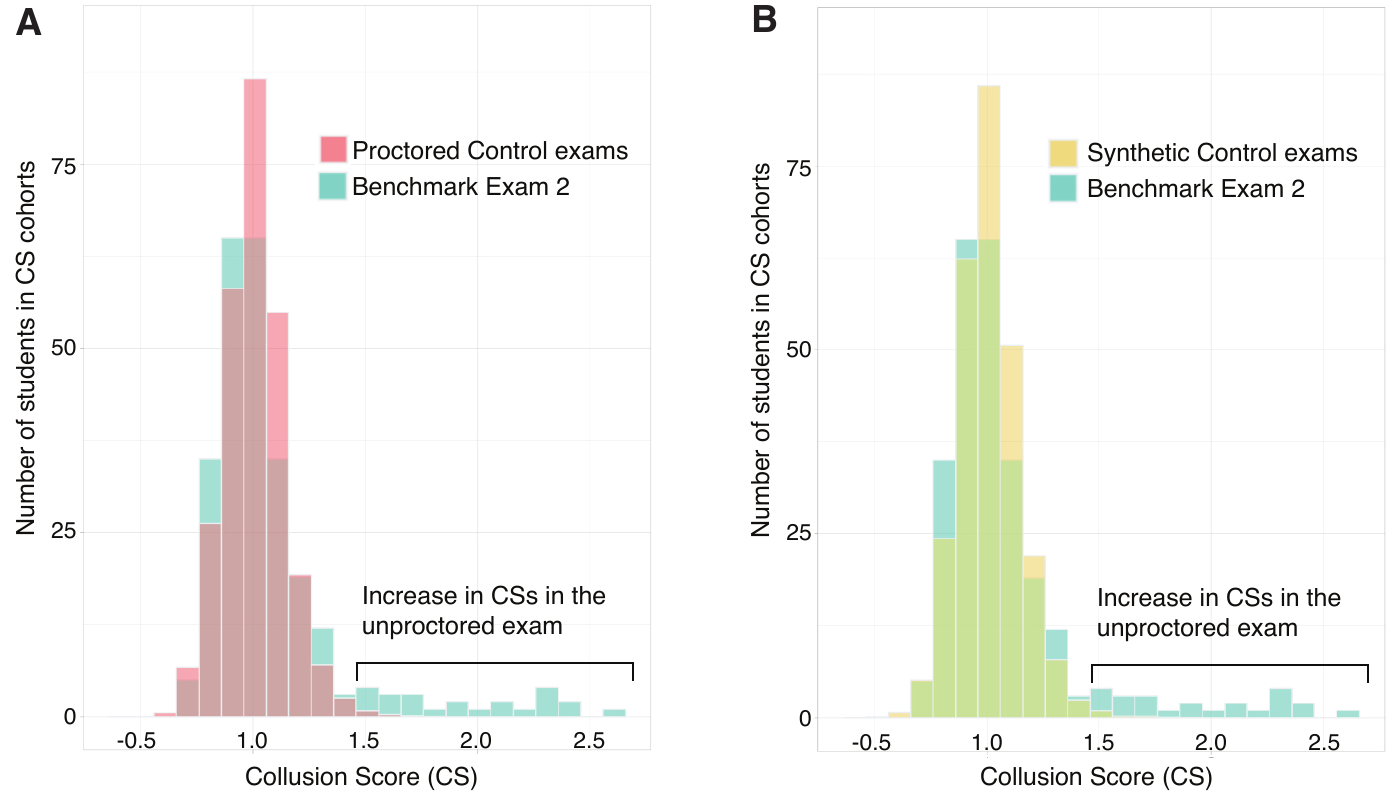}
\caption{The distributions of CS values from the $19$ proctored control exams (red), the unproctored benchmark exam 2 (blue), and the synthetic control exams (yellow). The distribution from the benchmark exam 2 has a heavier right tail than the other two distributions from control exams. \label{fig:CS_null}}
\end{figure}

Fig.~\ref{fig:CS_null}A compares the CS value distribution of the $6{,}847$ students from the $19$ proctored control exams to the distribution from the unproctored benchmark exam 2. As expected, the distribution from the unproctored benchmark exam 2 has a heavier right tail than that of distribution from the proctored control exams. 

\begin{figure}[htbp]
\centering
\includegraphics[width=0.65\textwidth]{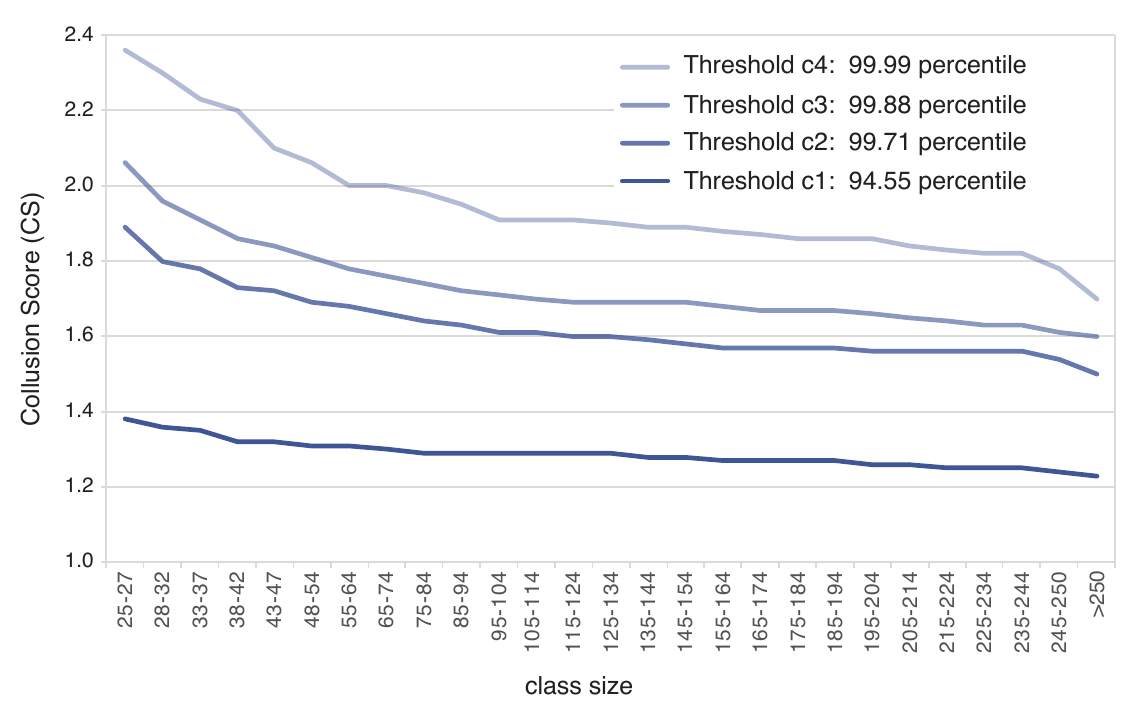}
\caption{CS value thresholds that correspond to the four CS distribution percentiles for varying class sizes subsampled from the proctored control exams.\label{fig:CS_thresh}}
\end{figure}

To help instructors decide which of the $M$ collusion groups, $\mathcal{G}_1, \ldots, \mathcal{G}_M$, detected in an unproctored exam are worth examining, we categorized the collusion groups into three \textit{bins} to reflect risk levels---\textit{low}, \textit{medium}, and \textit{high}---by thresholding the maxCS values of collusion groups. Specifically, we empirically define two maxCS thresholds, $c_3$ and $c_4$, based on the exam size, by downsampling the two benchmark exams, both containing more than $250$ students.

First, based on the two benchmark exams, we set $c_3=1.6$ and $c_4=1.7$ for exams with more than $250$ students. We chose the two maxCS thresholds so that the three bins,  containing collusion groups with maxCS values in $[c_2, c_3)$, $[c_3, c_4)$, and $[c_4, \infty)$, have similar numbers of collusion groups.

To account for the effects of exam size on maxCS values, similar to $c_1$ and $c_2$, we pre-calculated the thresholds $c_3$ and $c_4$ for an array of exam sizes: 15, 20, $\ldots$, 45, 50, 60, $\ldots$, 240, 250 by a subsampling-and-quantile-matching approach consisting of three steps. First, we converted the benchmark-exam-thresholds $c_3 = 1.60$ and $c_4 = 1.70$ to empirical null CDF values as $99.88\%$ and $99.99\%$ based on the empirical null distribution of CS values obtained from the $19$ proctored control exams (Table~\ref{tab:CS_thresh}). Second, for each exam size $n$ in $\{15, 20, \cdots, 45, 50, 60, \cdots, 240, 250\}$, we downsampled each of the $19$ proctored control exams to size $n$, calculated the CS value of each student within a downsampled exam, and repeat the downsampling and CS calculation for $100$ times. Then, we pooled the $(19 \times n \times 100)$ CS values to create an empirical null distribution for the exam size $n$. Third, we defined the $c_3$ and $c_4$ thresholds for the exam size $n$ as the $99.88\%$ and $99.99\%$ quantiles of the empirical null distribution (Fig.~\ref{fig:CS_thresh}).

Then, for exams with size no more than $250$, we rounded the exam size to the closest integer $n$ in the set of $\{15, 20, \cdots, 45, 50, 60, \cdots, 240, 250\}$. Then we used the $c_3$ and $c_4$ thresholds for exam size $n$ for this input exam to categorize the collusion groups into three bins.


With the categorization of the $M$ collusion groups into three bins, we calculated the empFPR for each bin from an expanded set of $36$ proctored exams taken by $n_e = 10{,}816$ students to increase the diversity of exams. We used the percent of these students (whom we believed to have not colluded)  placed into each bin to define the empFPR of that bin. Then, we obtained the empFPR values of the three bins as $0.04\%$, $0.20\%$, and $0.56\%$, corresponding to the low-, medium-, and high-risk levels. These empFPR values were defined to be cumulative: approximately $0.04\%$ of students in the $36$ proctored exams fell into the $0.04\%$ empFPR bin, $0.16\%$ fell into the $0.20\%$ empFPR bin, and $0.36\%$ fell into the $0.56\%$ empFPR bin. 

\begin{center}
\begin{table}[htbp]
\fontsize{10pt}{10pt}\selectfont
\centering
\caption{The CS percentiles used to define collusion groups and FPR bins. Each threshold is defined as a quantile of the CS values in the proctored control exams. All collusion group members must have CS values $\geq c_1$. All student/1st Partner pairs must have at least one member with a CS value $\geq c_2$. Thresholds $c_3$ and $c_4$ are used to assign collusion groups to three bins with different resk levels and empFPR values.  \label{tab:CS_thresh}}
\begin{tabular}{lp{3.5cm}p{3.6cm}p{5.9cm}}
  \toprule
  Threshold & CS value in the two benchmark exams & CDF value in the empirical null distribution of CS values & Interpretation \\ 
  \midrule
$c_1$ & $1.23$ & $94.55\%$ & Minimum CS value of detected colluded students \\ 
$c_2$ & $1.50$ & $99.71\%$ & Minimum maxCS value for the detected collusion groups in the high-risk bin ($0.56\%$ empFPR)\\ 
$c_3$ & $1.60$ & $99.88\%$ & Minimum maxCS value for the detected collusion groups in the medium-risk bin ($0.20\%$ empFPR)\\ 
$c_4$ & $1.70$ & $99.99\%$ & Minimum maxCS value for the detected collusion groups in the low-risk bin ($0.04\%$ empFPR).\\ 
   \bottomrule
\end{tabular}
\vspace{-.5cm}
\end{table}
\end{center}

We denoted $p$ to be the true FPR and assumed that $n$ students are independent Bernoulli random variables with probability $p$ to be mis-identified as a false positive. Then, we could write out the $95\%$ confidence interval of $p$ as $\widehat{p}_e\pm1.96\sqrt{\frac{\widehat{p}_e\left(1-\widehat{p}_e\right)}{n}}$, where $\widehat{p}_e$ denoting the empFPR was assumed to follow an approximately normal distribution with mean $p$ and variance $\frac{p\left(1-p\right)}{n}$. Table~\ref{tab:FPR_CI} summarizes the $95\%$ confidence intervals of true FPR for the three empFPR bins.



\begin{table}[htbp]
\fontsize{10pt}{10pt}\selectfont
\centering
\caption{$95\%$ confidence intervals (CIs) of true FPR for the three empFPR bins. \label{tab:FPR_CI}}
\begin{tabular}{|l|l|l|l|}
  \hline
empFPR bin & empFPR-based 95\% CI & \thead{synFPR-based 95\% CI \\ (Benchmark exam $1$)}   & \thead{synFPR-based 95\% CI \\ (Benchmark exam $2$)}\\
   \hline
  $0.04\%$ &$0.037\%~ (\pm 0.036\%)$ & $0.058\%~(\pm 0.001\%)$ &	$0.138\%~ (\pm 0.002\%)$\\
      \hline
   $0.20\%$ &  $0.203\%~ (\pm 0.085\%)$ & $0.195\%~ (\pm 0.002\%)$ &	$0.315\%~ (\pm 0.003\%)$\\
      \hline
   $0.56\%$ & $0.555\%~(\pm 0.140\%)$ & $0.623\%~ (\pm 0.003\%)$&	$0.697\%~ (\pm 0.003\%)$\\
      \hline
\end{tabular}
\end{table}

\subsection{Synthetic FPR}\label{subsec:synFPR}
Q-SID's empirical FPRs have proven useful. However, a potential source of error is that the proctored exams used to calculate them do not represent general exams, which are heterogeneous. In addition to class size, general exams vary in the number of questions asked, the difficulty of each question, the number of points awarded per question, and the distribution of the sum of each student's question scores (i.e., the distribution of a class' test scores). We, therefore, sought an additional estimate of the FPR based on the specific query exam submitted to Q-SID for analysis. To achieve this goal, we generated \textit{in silico} ``negative controls" of the query exam, in a way that the \textit{in silico} negative-control exams do not include any colluded students but still preserve the distributions of question scores and test scores in the query exam. We refer to such \textit{in silico} data as \textit{synthetic null data} or \textit{synthetic control exams}. 

\begin{figure}[htbp]
\centering
\includegraphics[width=0.8\textwidth]{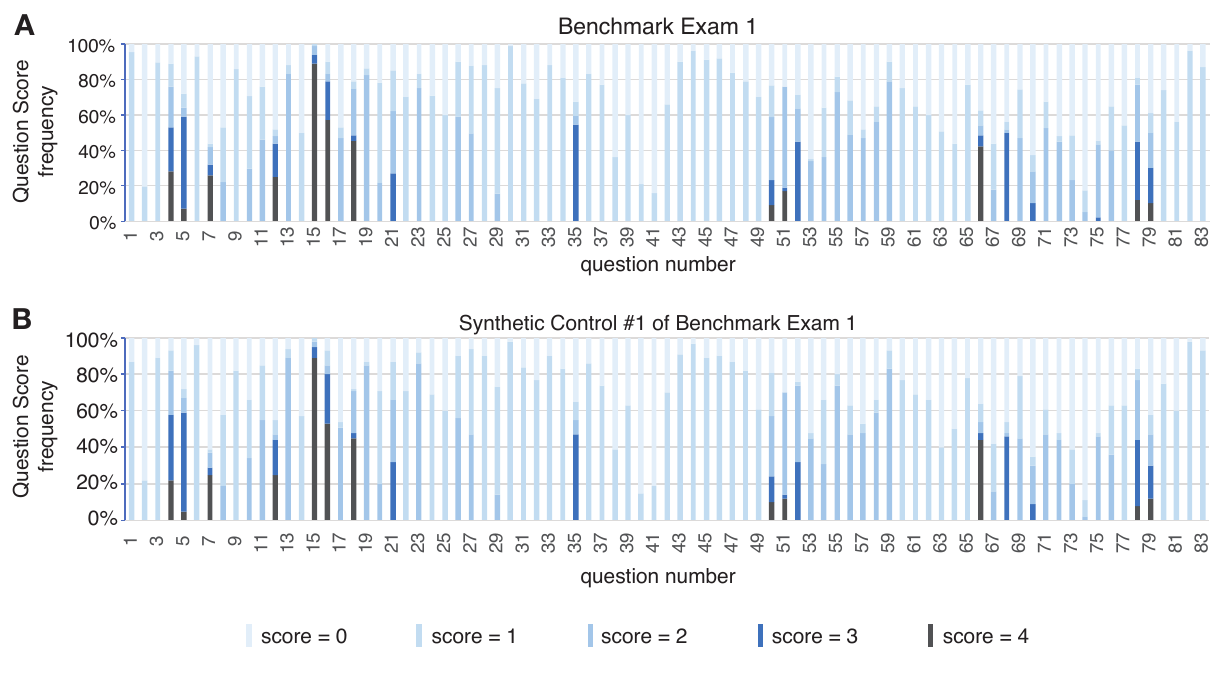}
\caption{The frequencies of scores awarded to students for each of the $83$ questions in (A) benchmark exam $1$ and (B) a synthetic control exam for benchmark exam $1$. The two distributions are similar, suggesting that the synthetic control exam preserves the question score distributions in benchmark exam 1. \label{fig:score}}
\end{figure}

Fig.~\ref{fig:score}A shows the frequencies of question scores in benchmark exam 1. The questions in this example have maximum scores between $1$ to $4$. Some questions are straightforward, so most students obtained the maximum score, e.g., question $1$. Other questions are more challenging, and only a minority of students obtained the maximum score, e.g., questions $78$ and $79$. Students who performed well on one challenging question tend to perform well on other challenging questions, and vice versa for those students who performed poorly. As a result, questions with similar difficulty levels tend to be positively correlated in terms of students' question scores. There are, thus, similarities among students' question scores that are not due to collusion but reflect different familiarities with the course material and the correlations among questions. A valid synthetic control exam must capture the correlations among questions  (Supplementary Material~\ref{subsec:supp_synthMimic}).

The statistical technique Gaussian copula can be employed to estimate the joint distribution of multiple variables—in our case questions—by coupling the estimated marginal distributions of these variables and preserving the rank correlations among these variables (\cite{nelsen2006introduction}). Previous work has shown that this technique can be used to generate realistic synthetic data in single-cell biology (\cite{sun2021scdesign2}). Here, we modeled the marginal distribution of every question (i.e., the distribution of the question scores across students) as a multinomial distribution, which agrees with the nature of question scores (every question has only a finite number of possible scores), and we leveraged the copula technique to estimate the joint distribution of all questions. The estimated joint distribution is assumed to be a null distribution that represents a homogeneous student population without collusion groups. We then sampled ``synthetic students" independently from this null distribution to generate synthetic control exams (Supplementary Material~\ref{subsec:supp_synthGenerate}). 

Crucially, we discovered that to generate synthetic controls whose distributions of test scores closely resemble those of a submitted real exam, it is necessary to fit the joint distribution of questions conditional on the test score (i.e., the sum of question scores for a student). To implement this strategy, we first divided students in the query exam into five groups based on the ranking of students' test scores, so that the students ranked in the top $20\%$ were placed in one group, the students ranked between the top $20\%$ and $40\%$ were placed in another group, etc. Then we fit a joint distribution of questions for each group and generated synthetic data by sampling ``synthetic students" from the fitted distribution. Finally, we pooled the ``synthetic students" from the five groups to form a synthetic control exam. 

The frequencies of question scores in synthetic control exams are similar to those in the corresponding real exams (e.g., Fig.~\ref{fig:score}). Synthetic control exams also closely mimic the means, variances, and coefficients of variation of questions in real exams, as well as the pairwise correlations among questions (Supplementary Material ~\ref{subsec:supp_synthMimic} and Fig.~\ref{fig:supp_SynReal}). 

We also verified that this generation approach produced synthetic control exams mimicking real exams of varying class sizes and question numbers. As a verification, we downsampled each of the 19 proctored control exams to two class sizes of $n=25$ and $200$, and for each exam and each class size $n$ we downsampled 50 times. Then for each downsampled exam of size $n$, we generated a total of $100{,}000/(50n)$ synthetic control exams. Finally, we calculated the median integration local inverse Simpson's index (miLISI) values between the downsampled exam and its corresponding synthetic control exams. For each of the $19$ proctored control exams and each $n$, we reported the average of the miLISI values (Supplementary Table 5).
A miLISI of $1$ indicates no similarity between a synthetic control exam and a real exam, while a miLISI of $2$ indicates a perfect similarity.  Across the $19$ proctored control exams, the mean of the average miLISI values are $1.967$ (SD $0.003$) and $1.956$ (SD $0.004$) for $n=25$ and $200$, respectively (Supplementary Table 5). Similarly, we downsampled the questions in the $19$ proctored control exams to $30$ or $60$ questions and carried out the synthetic exam generation. As a result, across the $19$ proctored control exams, the mean of the average miLISI values are $1.950$ (SD $0.004$) and $1.947$ (SD $0.008$) for $30$ and $60$ questions, respectively (Supplementary Table 6). These results indicate a high similarity between real and synthetic data, with no apparent differences due to the class size or question number.



	
For each query exam, Q-SID generates a sufficient number of synthetic control exams to include a total of $n_s>100{,}000$ synthetic students. For each synthetic control exam, CS values are calculated, and the synthetic students are placed into collusion groups and empFPR bins using the class-size specific CS thresholds established in the previous sections. 
Q-SID reports the cumulative proportions of students put in the three empFPR bins as the estimated \textit{synthetic FPR (synFPR)} $\widehat{p}_s$, and further calculates the $95\%$ confidence interval of the true FPR $p$ as $\widehat{p}_s\pm1.96\sqrt{\frac{\widehat{p}_s\left(1-\widehat{p}_s\right)}{n_s}}$, where $n_s$ denotes the number of synthetic students. 
For any query exam, while the three bins are set to have constant empFPRs (i.e., $0.04\%$, $0.20\%$, and $0.56\%$), the synFPRs are calculated based on synthetical control exams and thus not constants (e.g., Fig.~\ref{fig:supp_FPRs_table}). The $95\%$ confidence intervals of the true FPR for the two benchmark exams are summarized in Table~\ref{tab:FPR_CI}.

\subsection{Comparison between empirical and synthetic FPRs}
\begin{figure}[htbp]
\centering
\includegraphics[width=0.4\textwidth]{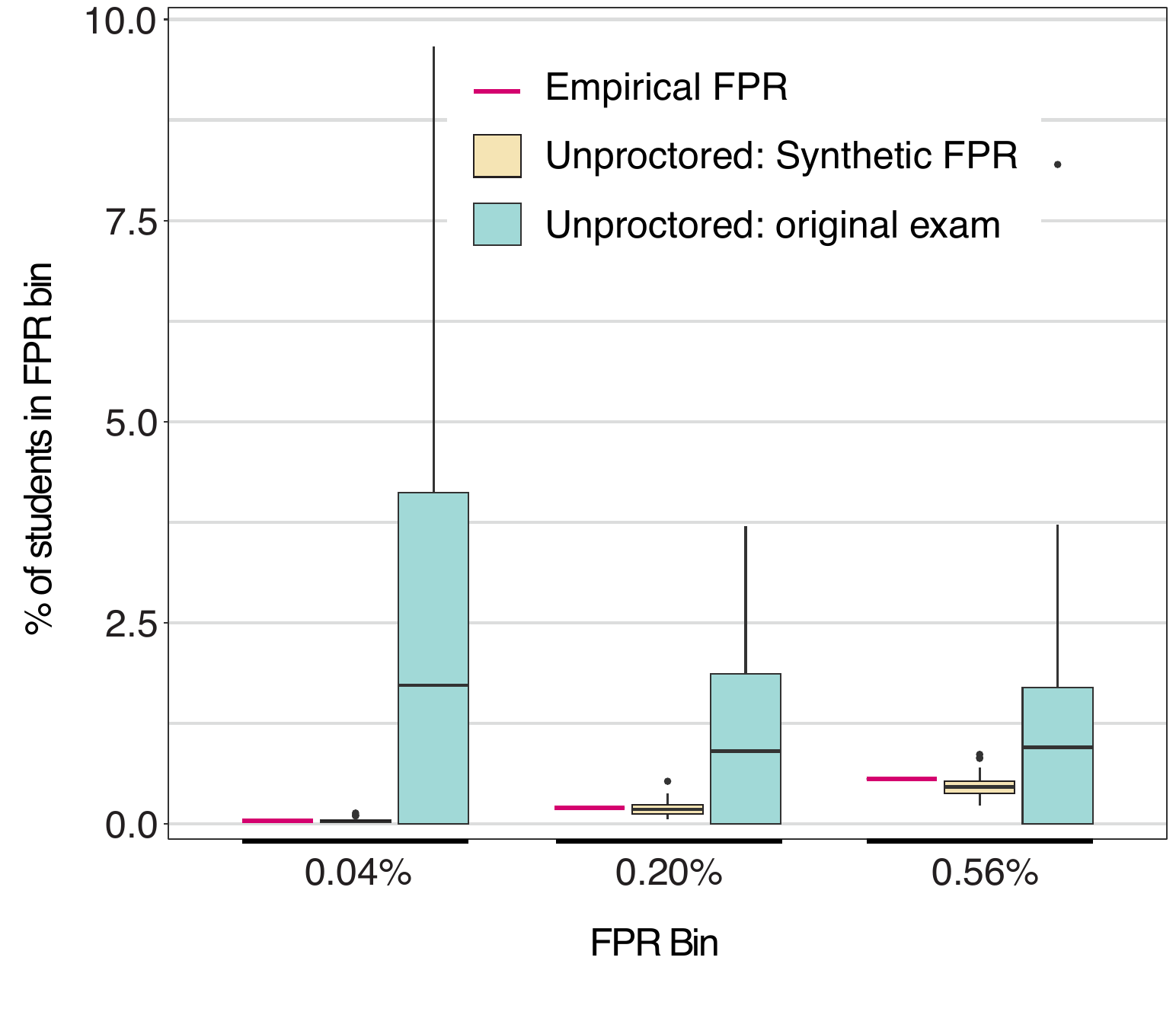}
\caption{The percents of students placed into each empFPR bin for the 34 unproctored exams (green) are shown together with the synthetic FPRs calculated for these same exams (brown). The empirical FPRs are shown by horizontal red lines.   \label{fig:FPRperc}}
\end{figure}

The distributions of CS values in the synthetic control exams of benchmark exam 2 closely resemble those in the 19 proctored controls, confirming the validity of the synthetic control exams (e.g., Fig.~\ref{fig:CS_null}, compare panels A and B). The similarity between empirical and synthetic FPRs suggests both are reasonable and trustworthy. Given this, when the proportion of students identified in each bin far exceeds these FPRs, collusion is the likely cause. Fig.~\ref{fig:FPRperc} shows that the percent of students flagged for collusion in the 32 unproctored exams significantly exceeded both empirical and synthetic FPRs. Instructor feedback confirms that many students identified by Q-SID were found to have colluded. Additionally, we confirmed the synthetic FPRs for many query exams aligned with the empirical FPR estimate, both being much lower than the percentage of students identified as colluding in unproctored exams (Fig.~\ref{fig:FPRperc}). Approximately half of the students flagged in unproctored exams fell into the $0.04\%$ empFPR bin with similar synFPRs. Higher identification rates in the $0.20\%$ and $0.56\%$ empFPR bins further supported collusion but with less clear discrimination than the $0.04\%$ empFPR bin (Fig.~\ref{fig:FPRperc}).


\begin{figure}[htbp]
\centering
\includegraphics[width=0.35\textwidth]{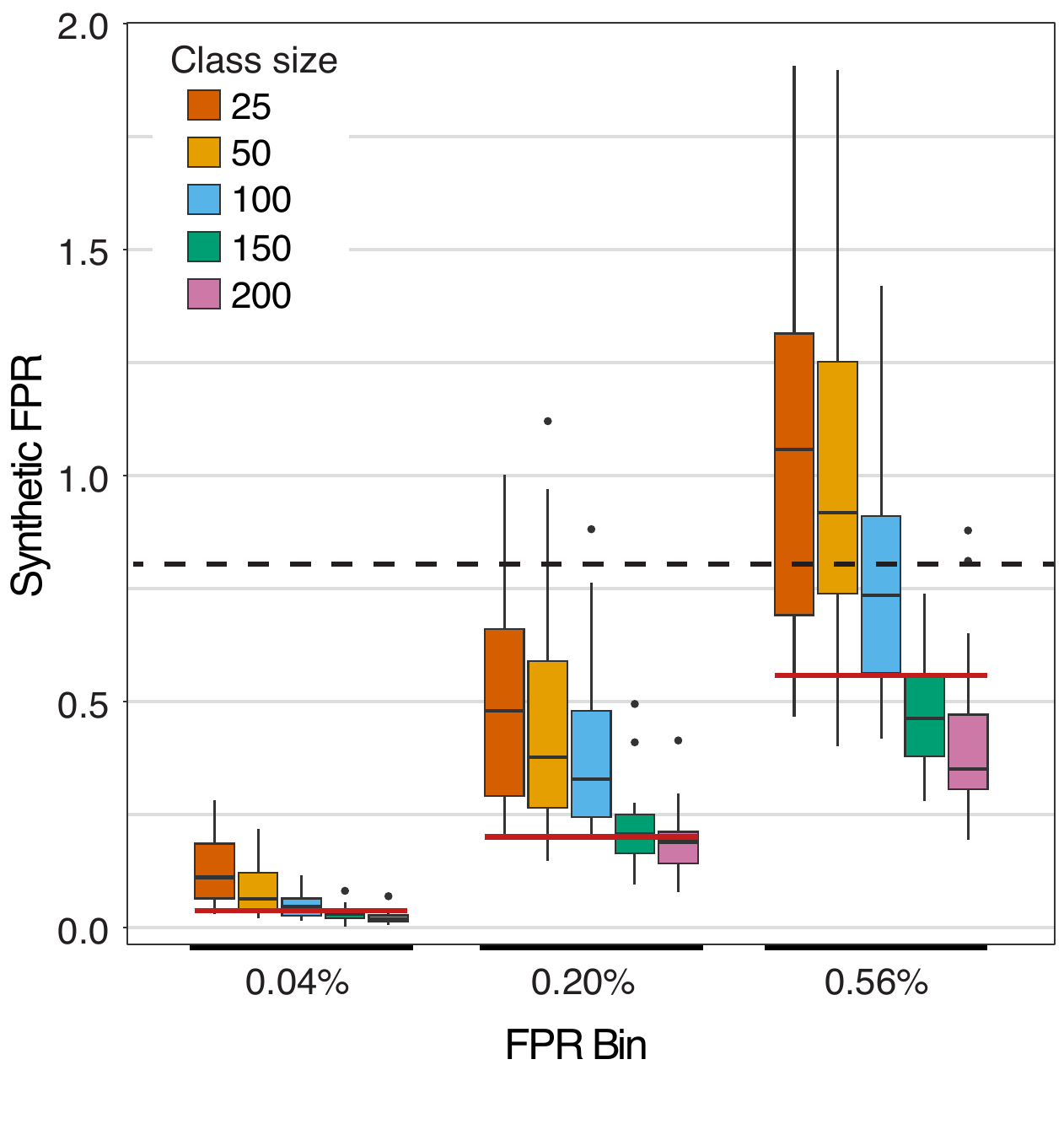}\label{fig:syntheticFPRsubsample}
\caption{The change in synthetic FPR with class size. Each of the $19$ proctored control exams were randomly down-sampled to produce 50 downsampled exams at each of the five class sizes indicated. Synthetic control exams were generated for sets of least 100,000 students for each exam size from each of the 19. For each set, the synthetic FPRs were determined and plotted as bar and whisker plots. The empirical FPRs are shown as horizontal red lines. \label{fig:FPRs}}
\end{figure}

To more systematically assess the agreement between Q-SID's empirical and synthetic FPRs, we generated randomly down-sampled exams from the $19$ proctored control exams that contained either different numbers of students or different numbers of questions and then calculated synFPRs. We found no systematic differences between synFPRs for downsampled exams that had different numbers of questions (Supplementary Material~\ref{subsec:supp_synFPR_compare}). However, there is a consistent difference between synFPRs for downsampled exams with different numbers of students (Fig.~\ref{fig:FPRs}). Exams with fewer than $100$ students tend to have synFPRs higher than the empFPRs (data above the horizontal red bars), whereas exams with more than $100$ students tend to have synFPRs lower than the empFPR (data below the horizontal red bars). 

The empirical and synthetic FPRs are equally valid. There is no evidence that shows one is more accurate than the other. Feedback from instructors, as well as our own inclination, strongly favor taking a conservative approach when the two FPRs diverge. As a precaution, therefore, Q-SID excludes any collusion groups whose cumulative synFPR is $>0.8\%$, i.e., any cases shown above the black dashed line in Fig.~\ref{fig:FPRs}. As a result, for exams with fewer than $100$ students Q-SID will often exclude collusion groups that would normally be placed in the $0.56\%$ empFPR bin, and in rare cases in the $0.20\%$ empFPR bin. As a result, fewer of the students who collude in small classes will be identified. However, in the $34$ unproctored exams, because three quarters of students who colluded were found in the $0.04\%$ and $0.20\%$ empFPR bins (Fig.~\ref{fig:FPRperc}), our results showed that Q-SID is still effective for the smaller exams within the $34$.

\begin{algorithm}[htbp]\label{alg:QSID} 
\textbf{Algorithm: Q-SID}
\footnotesize
\begin{tabbing}
   \enspace \textbf{Input:} a student-by-question score matrix $\mathbf{X} \in \mathbb{R}_{\ge 0}^{n \times p}$, with $n$ students and $p$ questions\\
   \enspace \textbf{Internal parameters:} \\
   \qquad\qquad Collusion score (CS) thresholds $0 < c_{1} < c_{2} < c_{3} < c_{4}$ given the class size $n$; \\
    \qquad\qquad Empirical FPR  levels $f_1 = 0.04\%$, $f_2=0.20\%$, and $f_3=0.56\%$.\\
    
    
    \enspace  \textbf{Step 1 (Colusion scores and partners).} For each student $i=1,\ldots, n$, \\
    \qquad\qquad \textbf{1.} Calculate $\text{IS}_{ij}$ for every other student $j \neq i$.\\
     \qquad\qquad \textbf{2.} Calculate $\text{IM}_i=\underset{j\neq i}{\max} {{\,} \text{IS}_{ij}} - \underset{j\neq i}{\text{median}}{\,}{\text{IS}_{ij}}$.\\
    \qquad\qquad \textbf{3.} Calculate $\text{CS}_i=\text{IM}_i/\left(\text{local\ median\ IM}_i\right)$, where the local median $\text{IM}_i$ is the \\
    \qquad\qquad\qquad median IM of student $i$'s neighbors (defined based on the test score rank).\\
    \qquad\qquad \textbf{4.} Identify student $i$'s 1st partner $i_1 = \underset{j \neq i}{\argmax} {\,} \text{IS}_{ij}$ and 2nd partner $i_2 = \underset{j \neq i, i_1}{\argmax}{\,} \text{IS}_{ij}$.\\

    \enspace  \textbf{Step 2 (Provisional collusion groups).} Identify the $\{$student, 1st partner$\}$ sets $\{i, i_1\}$ \\
   \qquad\qquad such that $\text{CS}_i \ge c_1$, $\text{CS}_{i_1} \ge c_1$, and at least one of $\text{CS}_i$ and $\text{CS}_{i_1}$ is no less than $c_2$,\\
    \qquad\qquad   with $i \neq i_1 \in \{1, \ldots, n\}$. Combine the  overlapped sets into $K$ non-overlapping \\
    \qquad\qquad    provisional collusion groups $\mathcal{T}_1, \ldots, \mathcal{T}_K$, each of which is a set of students. \\
    \qquad\qquad  The $K$ groups are ordered by the maximum CS of the students in each group, in \\
    \qquad\qquad a decreasing order.\\
    \enspace  \textbf{Step 3 (Collusion groups).} Detect collusion groups  $\mathcal{G}_1, \ldots, \mathcal{G}_M$, where $M \leq K$. Initialize $m=1$.\\
    \qquad\qquad For $k = 1,\ldots,K$,\\
    \qquad\qquad - If $\mathcal{T}_k \neq \emptyset$:\\
    \qquad\qquad\qquad \textbf{1.}    Initialize $\mathcal{G}_m=\mathcal{T}_k$.\\
    \qquad\qquad\qquad  \textbf{2.}  Identify the duplicate 2nd partner set \\
    \qquad\qquad\qquad\qquad  $\mathcal{P}_k = \{l \in \{1,\ldots,n\} \setminus \mathcal{T}_k :\exists i \neq j \in \mathcal{T}_k \, \text{ s.t. }{i_2}={j_2}=l \}$, where $i_2$ and $j_2$ are \\
    \qquad\qquad\qquad\qquad   students $i$ and $j$'s 2nd partners, respectively. \\
    \qquad\qquad\qquad  \textbf{3.}  For each student $l$ in $\mathcal{P}_k$, if $\exists ~ \mathcal{T}_s,~s > k$, s.t. $l \in \mathcal{T}_s$:\\
    \qquad\qquad\qquad\qquad  - update $\mathcal{G}_m = \mathcal{G}_m \cup \mathcal{T}_s$,\\
    \qquad\qquad\qquad\qquad  - update $\mathcal{T}_s = \emptyset$.\\
    \qquad\qquad\qquad \textbf{4.}    Update $m=m + 1$.\\
    \qquad\qquad  - Else if $\mathcal{T}_k =\emptyset$, skip.\\
    
\end{tabbing}
\end{algorithm}

\begin{algorithm}[htbp]\label{alg:QSID2} 
\footnotesize
\begin{tabbing}
 \enspace  \textbf{Step 4 (Empirical FPRs).} For each collusion group $\mathcal{G}_m, m=1,\ldots,M$, denote its  \\
    \qquad\qquad  maximum group member CS value as $\text{CS}^*_{m} = \underset{i \in \mathcal{G}_m}{\max} {\,}\text{CS}_i$, which is at least $c_2$ \\
   \qquad\qquad   by the definition of $\mathcal{G}_m$. \\ 
    \qquad\qquad \textbf{1.}  If $\text{CS}^*_{m} \geq c_4$, assign $\mathcal{G}_m$ with the empirical FPR $\text{empFPR}_m = f_1$. \\
    \qquad\qquad \textbf{2.}  If $c_3 \leq \text{CS}^*_{m} < c_4$, assign $\mathcal{G}_m$ with the empirical FPR $\text{empFPR}_m = f_2$.\\
    \qquad\qquad \textbf{3.}  If $c_2 \leq \text{CS}^*_{m} < c_3$, assign $\mathcal{G}_m$ with the empirical FPR $\text{empFPR}_m = f_3$.\\
    
    \enspace  \textbf{Step 5 (Synthetic FPRs: calibration of empirical FPRs).} \\
    \qquad\qquad \textbf{1.} Simulate a synthetic null student-by-question score matrix $\tilde{\textbf{X}}  \in \mathbb{R}_{\ge 0}^{n \times p}$ by \\
     \qquad\qquad\qquad fitting a null model on $\textbf{X}$. In the null model, no students colluded. \\
    \qquad\qquad \textbf{2.} Determine the synthetic null collusion group $\tilde{\mathcal{G}}_m$, $m=1,\ldots,\tilde{M}$, and their  \\
    \qquad\qquad\qquad  corresponding empirical FPRs $\in \{f_1, f_2, f_3\}$ using the synthetic null data $\tilde{\textbf{X}}$ \\
    \qquad\qquad\qquad by following Step $1$ to Step $4$. Then each synthetic null student in any of \\
    \qquad\qquad\qquad $\tilde{\mathcal{G}}_1, \ldots, \tilde{\mathcal{G}}_{\tilde M}$ is assigned with an empirical FPR of $f_1$, $f_2$, or $f_3$.\\
    \qquad\qquad \textbf{3.} For each of $l=1,2,3$, calculate the proportion of synthetic null students that\\
    \qquad\qquad\qquad  are assigned  with the empirical FPR level $f_l$; denoted the proportion as $\tilde{f}_l$.\\
    \qquad\qquad \textbf{4.} For each real collusion group $\mathcal{G}_m, m=1,\ldots,M$,\\
     \qquad\qquad\qquad \textbf{(1)} if $\text{CS}^*_{m} \geq c_4$, assign $\mathcal{G}_m$ with the synthetic FPR ${\text{synFPR}}_m = \tilde{f}_1$;\\
     \qquad\qquad\qquad \textbf{(2)} if $c_3 \leq \text{CS}^*_{m} < c_4$, assign $\mathcal{G}_m$ with the synthetic FPR ${\text{synFPR}}_m = \tilde{f}_2$;\\
     \qquad\qquad\qquad \textbf{(3)} if $c_2 \leq \text{CS}^*_{m} < c_3$, assign $\mathcal{G}_m$ with the synthetic FPR ${\text{synFPR}}_m = \tilde{f}_3$.\\
    
    \enspace \textbf{Output:} Student collusion groups $\mathcal{G}_1, \ldots, \mathcal{G}_M$ with their corresponding empirical FPRs \\
    \enspace \qquad\qquad ${\text{empFPR}}_1, \ldots, {\text{empFPR}}_M$ and synthetic FPRs ${\text{synFPR}}_1, \ldots, {\text{synFPR}}_M$.
\end{tabbing}
\end{algorithm}

\section{Power assessment of Q-SID}\label{sec:TPR}
For a collusion detection method to be useful, its effectiveness must be shown across a wide array of exam types and formats. We therefore systematically measured the percent of true positives placed into collusion groups, i.e., the \textit{true positive rate (TPR)}, also known as the \textit{power}. Of necessity, our study uses the two benchmark exams as only for these has a set of students been rigorously shown to have colluded by analysis of written answers and confessions. It is likely that this true positive set represents nearly all of the students who colluded on these exams. This conclusion is supported by the fact that an analysis of the written answers of numerous other students, whose question scores were also highly correlated, failed to detect any instances of collusion (Section \ref{subsec:data_benchmark}). Thus, it is reasonable to use these data to determine the TPR. We have used subsampling or combining exams to systematically vary and test the power of Q-SID under varying question numbers, question score complexities, and class sizes.

\begin{figure}[htbp]
\centering
\includegraphics[width=\textwidth]{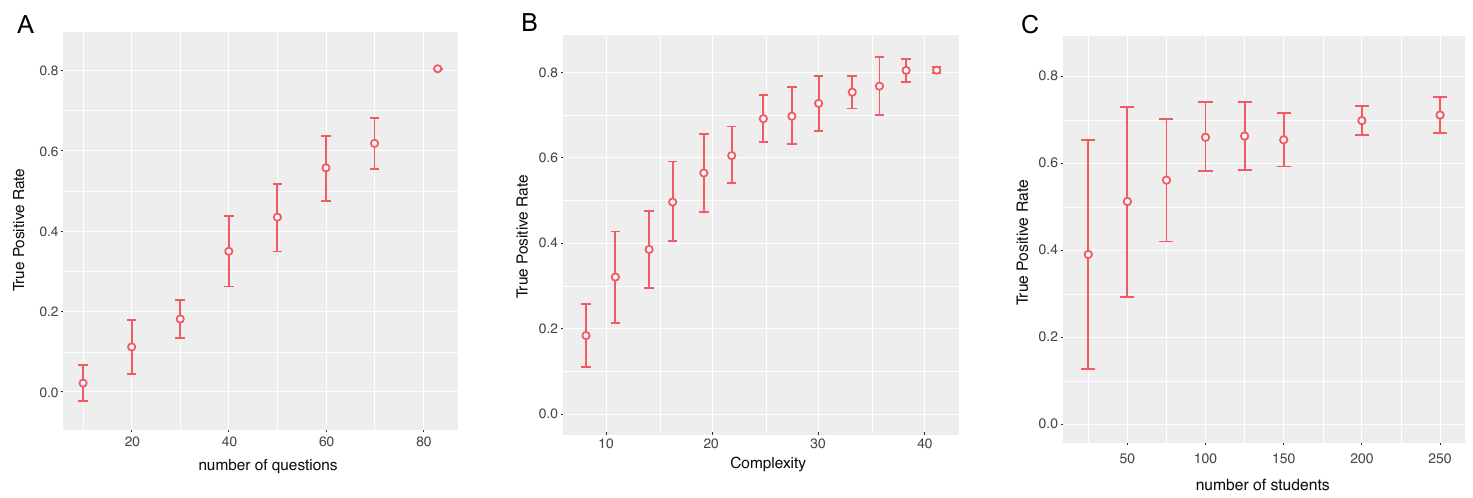}
\caption{(A) The TPR for varying numbers of questions per exam. (B) The TPR for exams with varying complexities. (C) The TPR for exams with varying numbers of students. \label{fig:TPR}}
\end{figure}

\subsection{Number of questions}\label{subsec:TPR_question}

Q-SID performance is strongly influenced by the number of questions. To analyze this, numeric scores for randomly sampled subsets of questions were drawn from benchmark exam 1 to create virtual exams. Fig.~\ref{fig:TPR}A shows that, on average, only $10\%$ of true positives are identified in downsampled exams with $20$ questions, while $55\%$ are found in exams with $60$ questions, and $80\%$ with $83$ questions. Additionally, the distribution of CS values becomes less reliable in virtual exams with fewer than $20$ questions. Therefore, Q-SID does not process exams with fewer than $20$ questions.



\subsection{Question score complexity}\label{subsec:TPR_comp}
In addition to being affected by the number of questions, Q-SID performance is influenced by the frequency with which a class obtains similar or different scores on each question. Instructors differ in their exam styles. At one extreme, a confirmatory exam may be set where most questions are answered correctly by nearly all students, resulting in many identical scores. Conversely, in a more rigorous exam, students may obtain a wide range of scores for each question. Questions that provide the greatest discrimination between students are more powerful for detecting collusion. For example, analysis of the $50$ questions with the smallest variation in scores identifies only $7\%$ of true positives in the two benchmark exams. In contrast, analysis of the 50 questions with the largest variation identifies $56\%$ of true positives. 
	
To account for both the number of questions and the discriminatory power of each question, Q-SID calculates a value called \textit{complexity} for each exam as follows. Complexity $O_s$ for question $s=1,\ldots,p$ is defined as
\begin{eqnarray*}
O_s=-{\log}_{10}{\left(\sum_{x\in\mathcal{X}}{\widehat{p}_s\left(x\right)^2}\right)}\,,
\end{eqnarray*}
where $\widehat{p}_s\left(x\right)=\sum_{i=1}^n1\{{X}_{is}=x\}/n$ is the relative frequency of the observed numeric score $x$, and $\mathcal{X}$ is the finite sample space of scores for question $s$. We further define the complexity $O$ for an exam as 
\begin{eqnarray*}
O=\sum_{s=1}^p O_s\,.
\end{eqnarray*}

The question score data from the two benchmark exams were first combined, and then varying numbers of questions were downsampled to produce exams with different levels of complexity. Fig.~\ref{fig:TPR}B shows that the true positive rate (TPR) increases with complexity, following an approximately hyperbolic trend. The steepest increases in TPR occur for complexities up to $20$.

Complexity measures the amount of information present in question score data. Unlike the number of questions, complexity should be a universal metric: exams with the same complexity should yield similar TPRs, except for classes with fewer than $100$ students.

It is not necessary to detect every student who colludes in every exam since students are usually assessed by multiple exams during a term. To efficiently detect collusion, Fig.~\ref{fig:TPR}B suggests arranging exams into sets with complexities between $15$ and $25$. If the complexities of two or more exams are below $15$, their question scores should be combined and analyzed together. The complexity of a combined exam is calculated by summing the complexities of the individual exams. Analyzing several exams with complexities of $15$ or higher should detect about $80\%$ of students who colluded (Fig.~\ref{fig:TPR}B).

To predict the likely complexity of a new exam and the expected TPR, instructors can use Q-SID to analyze prior exams with similar numbers and styles of questions. In our dataset of $70$ exams, the mean complexity per question is $0.33$, with a minimum of $0.16$ and a maximum of $0.65$.

\subsection{Number of students}

Randomly selected subsamples of 25 to 250 students were used to model different class sizes. Fig.~\ref{fig:TPR}C shows that there is little variation in the TPR for classes with $100$ or more students. For classes with fewer than $100$ students, the mean TPR decreases progressively while the standard deviation increases. The reduction in TPR for classes with fewer than $100$ students is primarily due to the exclusion of collusion groups with a synthetic FPR greater than $0.8\%$. Without this exclusion step, the mean TPR remains relatively constant across all class sizes (data not shown). We have set the lower limit for the number of students that Q-SID will analyze at $25$ because the variation in performance is unacceptable for classes smaller than this.


\section{Collusion Detection with Q-SID}\label{sec:det_col}



Although Q-SID was initially developed to identify students who had colluded, it soon became evident that preventing cheating before exams was more effective. Reminding students of the honor code and having them sign a statement pledging not to cheat failed to prevent collusion in the Benchmark and other unproctored exams (Fig.~\ref{fig:FPRperc}). Consequently, we aimed to inform students taking online exams that we were using an algorithm highly effective at detecting collusion. Students were either given a link to a website with detailed information about Q-SID or informed that others in prior classes had been caught cheating using Q-SID.

\begin{figure}[htbp]
\centering
\includegraphics[width=0.4\textwidth]{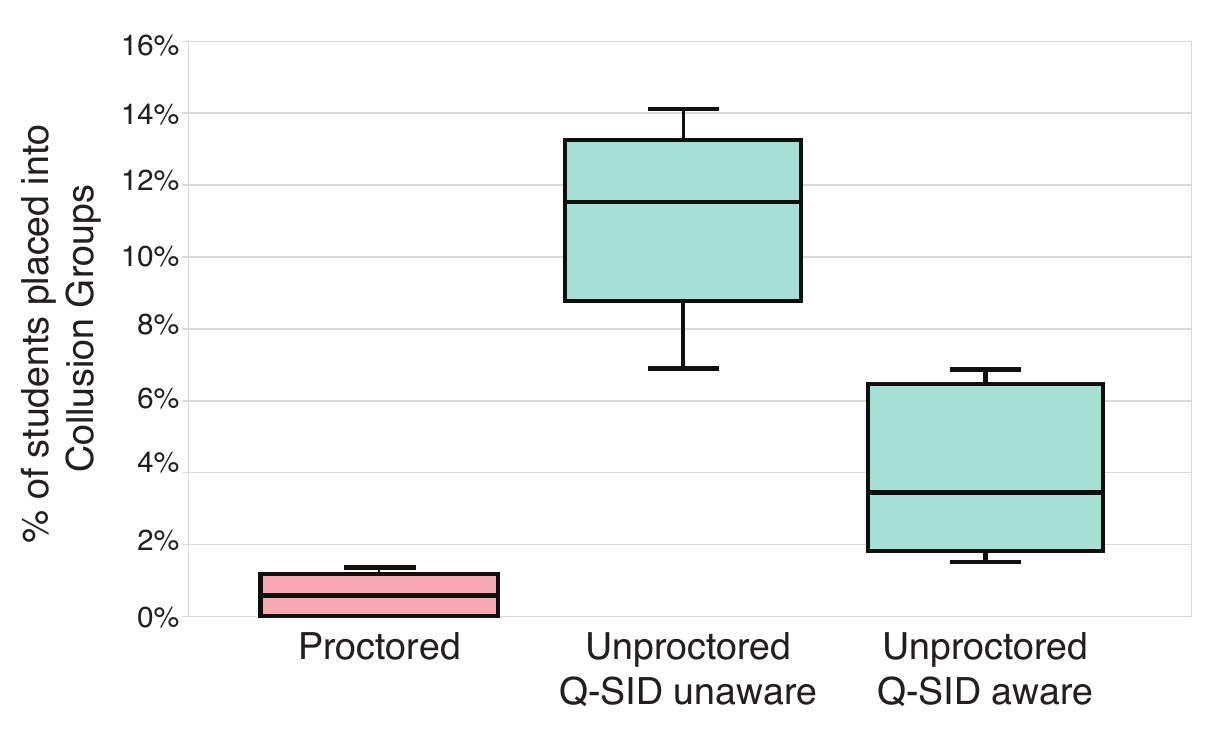}
\caption{The percent of students placed into collusion groups for exams taken under the circumstances shown. \label{fig:det_col}}
\end{figure}

Fig.~\ref{fig:det_col} illustrates the percentage of students assigned to collusion groups in exams from the same UC Berkeley course. For twenty-two exams administered in person with strict proctoring in the three years before the COVID-19 pandemic, few students were placed into collusion groups. For the first five online exams during the pandemic, students were unaware that Q-SID would be used to detect collusion, resulting in $7\%$ to $14\%$ of the class colluding. In the subsequent ten unproctored exams, students were either shown a website describing Q-SID or informed that peers had been caught cheating using Q-SID, leading to only $2\%$ to $7\%$ of students being identified as colluding. This small study suggests that informing students about Q-SID significantly reduced collusion.

Additionally, many students who colluded despite being informed about Q-SID confessed when confronted and indicated they would not cheat in the future. While these cases are not ideal, deterrence through sanctions is another way Q-SID discourages students from colluding in future classes.
\vspace{-.2cm}

\section{An Online Webtool}\label{sec:web}

To facilitate the use of Q-SID by instructors, we have made available a free public web tool \url{http://shiny2.stat.ucla.edu/Q-SID/} (Note to reviewers: this link leads to a non-public version of our web tool, which has been edited to remove author-identifying material). Instructors can upload a file containing question and test score data in the appropriate format. The website processes this data and returns a detailed report within two minutes. This report lists the members of collusion groups, their CS values, empirical FPRs, synthetic FPRs, and additional relevant information. An example report is provided in Supplementary Material~\ref{subsec:supp_report}.

\section{Discussion}\label{sec:disc}
We have demonstrated that Question-Score Identity Detection (Q-SID) can identify up to 80\% of students who have colluded on an exam with false positive rates (FPRs) ranging from 0.04\% to 0.8\%. Q-SID is unique in its ability to detect collusion in both multiple-choice and non-multiple-choice exams, provide two independent estimates of the FPR (empirical and synthetic), and predict the true positive rate (TPR) for various exam formats. In an analysis of 34 unproctored online exams taken during the COVID-19 pandemic by 10,526 students, Q-SID identified 4.5\% of examinees as likely colluders, with 2.4\% detected at an empirical FPR of 0.04\% and similarly low synthetic FPRs (Fig. \ref{fig:FPRperc}).

For an exam with 200 students, there are 19,800 possible pairwise combinations of exams whose written answers would need to be compared to detect all cases of collusion. This task is impossible to perform manually, and current text analysis algorithms cannot distinguish copied answers from those that are similar because they are correct. Q-SID addresses this problem by highlighting small groups of students who are likely, but not certain, to have colluded. Instructors can then compare the written answers of these students to determine if they are more similar to each other's than to those of other class members.

In addition to identifying specific students for further investigation, Q-SID can rapidly measure collusion across many classes without charging individual students with cheating. Because Q-SID's FPRs are calibrated and consistent between empirical and synthetic FPRs, the algorithm can reliably estimate the degree of collusion in each exam. We have found that some courses consistently exhibit higher levels of collusion, often reflecting students' perceptions of the course's importance. For example, in a course required for entry to many medical schools, a median of 6.5\% of students were identified as likely colluders across 15 unproctored online exams. Other exams showed much lower levels of collusion or none. Identifying which courses have higher levels of cheating can help instructors and administrators focus their efforts to address this problem.

Q-SID can also be used as a deterrent. The percentage of students who collude can be reduced two to threefold if they are informed about Q-SID before the exam (Fig. \ref{fig:det_col}). We believe that widespread adoption of Q-SID by institutions could largely eliminate collusion in online exams over time, as students would understand that they will be caught.

While Q-SID can analyze multiple-choice exams, it should only be used to gauge whether collusion has occurred, not to identify and challenge specific students for cheating. The lack of written answers prevents the additional analysis needed to establish compelling evidence of individual cheating. When collusion is detected in multiple-choice exams, instructors should consider alternative exam formats or improved proctoring in the future.

Cheating on exams has a corrosive effect. Students who cheat often receive higher grades than they deserve, while those who do not cheat may receive lower grades. The opportunity to cheat is greater in online exams than in traditional in-person exams due to the challenges of online proctoring. The increased use of online exams creates a profound need for methods to detect, measure, and deter collusion. Q-SID is a well-justified approach that addresses this need.

\section{Acknowledgements}\label{sec:ackn}
We are grateful to various instructors at UC Berkeley and UCLA who provided question score data from their classes as well as feedback on the effectiveness of Q-SID in identifying collusion. 

\section{Funding}\label{sec:fund}
This work was funded in part by a Sloan Research Fellowship and a Johnson and Johnson WiSTEM2D Scholars Award to JJL. 

\section{Supplementary Material}\label{sec:supp}
The article comes with six Supplementary Tables (Supplementary Tables 1--3 as Excel files, and Supplementary Table 4 as a pdf file, Supplementary Table 5--6 as csv files), a Supplementary Material pdf file, a zip file containing the R-code, and a zip file of Supplementary Data. Supplementary Table 1 is a summary of the question score data and Q-SID analysis for each exam. Supplementary Tables 2--3 provide the true positive and true negative labels for two benchmark exams, along with the CS values of students and their 1st partners. Supplementary Table 4 displays the class-size-specific CS thresholds for forming collusion groups and three empFPR bins. Supplementary Tables 5--6 summarize the mean miLISI values between downsampled exams and corresponding synthetic exams for the $19$ proctored control exams.
The Supplementary Material pdf contains additional details on the method. The R-code is released under the {Q-SID academic license}. The Supplementary Data contains all of the question score data used in this work together with lists of the true positives and true negatives.  All data is anonymous as to institution, department, course and student.


\appendix
\renewcommand\thefigure{S\arabic{figure}}    
\setcounter{figure}{0}    
\renewcommand\thetable{S\arabic{table}}    
\setcounter{table}{0}    


\newpage
\pagenumbering{arabic}

\begin{center}
{\large\bf SUPPLEMENTARY MATERIAL}
\end{center}

\section{Methodology}\label{sec:proofs}

\subsection{Manual comparison of written answers \label{subsec:supp_written}}
To determine if two or more students have colluded with sufficient certainty to charge them, the classic approach of manually comparing their written answers must be used. Collusion can be quickly ruled out for students who have not cheated. In our experience, a lack of similarity in detail in their answers is readily apparent after only a few minutes comparing the group's exams. For collusion groups where answers appear overly similar on a first pass through the exam, a more time consuming process is then required that can take several hours. This involves first finding unlikely or unusual aspects of the suspects answers to specific questions, then looking at all other students' answers. Where students have colluded, at least several examples can be found for which the only students giving a particular answer are the members of that group.

Certain questions are more dispositive of collusion than other. Within an instructors' style of question, those that elicit the greatest range of answers among students are typically those for which collusion can most clearly be demonstrated. Short answer questions for which the correct answer closely follows the lecture notes will likely be very similar for many members of the class, most of whom did not collude.

Among the 263 students who took both benchmark exam datasets from two unproctored, online exams in a UC Berkeley class during the pandemic, initially, many pairs of students were identified whose scores for each question had high Pearson correlation coefficients. $50$ of these students were subsequently found to have colluded by a time-consuming, manual analysis of written answers. The $50$ colluded in $19$ groups, comprised of between two to five members. $46$ of them were identified within the time allowed for formal charges to be made, and $37$ of these ($80\%$) subsequently confessed to collusion. The written answers of the nine students who did not confess contained as many compelling instances of collusion as the $37$ who did confess and the four who went uncharged. All $50$ students identified by textural analysis of answers are considered to be true positives.

As part of the month long process to identify the true positives, the written answers of an additional $26$ pairs of students whose question scores correlated highly were also compared and found to show no evidence of collusion. This set of $52$ students are designated as true negatives. There are $>30{,}000$ additional pair-wise combinations of student exams for this class whose written answers could not be compared manually. The Pearson correlation coefficients between these other pairs' question scores were generally lower than those of the true positives or true negatives, and thus these other pairs are unlikely to have colluded. These other students are not included in the true negative set and instead are referred to as uncharacterized students.
\subsection{Median Identity Metrics}\label{subsec:supp_medIM}
The median Identity Metric (median IM) for each student is calculated as follow. Students are first ranked by test score bringing students of similar ability into proximity. A sliding window that spans a set of adjacent students is used to produce local median values for IMs at every position in the test score rank list. For most students, the windows are centered on that student and extend an equal number of students above and below the student on the rank list. However, for the three students at each end of the rank list, the windows cannot be centered and instead the three students all share the same window. For classes of $31$ or more students, the windows cover $31$ students, except at the ends of the rank list, where they progressively shorten to $7$ students. As an example, for a student test score rank $50$ in a class of $100$, the median of the IMs for students ranks $35$ to $65$ is calculated. 

\subsection{Elimination of occasional error in calculating CSs}\label{subsec:supp_CSerrs}
In addition to the test score rank bias and the student specific bias, we have found two other sources of error that affect a small number of students in a minority of exams. 
	
The first of these sources of error results from students who score zero on most questions. Pairs of such students have aberrantly high IMs, which impacts the local median IM and can lead to biased CSs for students at the very bottom of the test score rank list. This source of error is removed by the simple expedient of excluding students whose test scores are $\le 5\%$ of the highest test score obtained in the exam. Students who scores are so extremely low are unlikely to have colluded and thus it is reasonable to remove them from the calculation of CSs. Frequently these students are later found to drop the class when they realize how poorly they have performed. 
	
The second source of error results from clerical mistakes that lead to data for one or more students being present more than once, often with similar but not identical question scores. If these data remained, such duplicate entries would lead to student/1st partner pairs with high CSs and would as a result be placed in collusion groups. To prevent this, Q-SID ignores data from any rows that share the same Student ID as well as any row lacking an ID. Q-SID list the IDs of any data it ignores in the Q-SID report pdf (see Supplementary Material~\ref{subsec:supp_report}).

\subsection{Class-size specific CS thresholds}\label{subsec:supp_CS}
The CSs plotted in Fig.~\ref{fig:CS_thresh} are shown in the lookup table (Supplemntary Table 4). These values are those employed by Q-SID to determine the absolute CSs for placing students into collusion groups and FPR bins based on the size of the class.

\clearpage
\section{Synthetic question score data}\label{sec:supp_synth}
\subsection{Generation of synthetic null data}\label{subsec:supp_synthGenerate}
Synthetic control exams are in silico synthetic versions of the query exam that capture the distributions of question scores and test scores, but do so in a way that does not include any correlations within the data due to collusion. To generate a realistic synthetic question score matrix, we include two steps: model fitting on the real question score matrix and generation of a synthetic question score matrix by sampling from the fitted model. Note that the marginal multinomial distribution requires integer values and many question scores are not integer, for example 1.5 or 0.33. Hence before fitting the model, we first multiply all scores by 100 then round to create integers that distinguish question scores that differ by $\geq0.01$. Then after we fit the model and generate the synthetic score matrix, we recover the simulated score matrix back to the original scale by dividing $100$.

We denote the integer question score matrix by $\mathbf{X}\in\mathbb{N}^{n\times p}$ with $p$ questions and $n$ students. We further assume that the $n$ students belong to $K$ cohorts, which are uniformly separated based on the student test scores’ quantiles (specifically, we use $K=5$ in our analysis), and we perform the simulation independently for each student cohort. We denote the $\mathbf{X}^{\left(k\right)}\in\mathbb{N}^{n^{\left(k\right)}\times p}$ as the question score sub-matrix corresponding to the student cohort $k$, where $n^{\left(k\right)}$ is the number of students belonging to cohort $k$. To simplify the notations, we omit the superscript $\left(k\right)$ for the $k$-th cohort in the following discussion of generative models.

\subsubsection{Fitting a generative model of the question score matrix}
The first step is to fit a generative model to characterize the joint distribution of questions’ scores across students for each student cohort. The model treats the questions as random variables and students as the independent and identically distributed (i.i.d.) observations. Here we model the marginal distribution of every question (i.e., the distribution of the question's scores across students) as a multinomial distribution, which agrees with the nature of question scores (every question only has a finite number of possible scores), and we leverage the copula technique to estimate the joint distribution of all questions. 

In the following analysis, we denote the question score matrix by $\mathbf{X}\in \mathbb{R}_{\geq0}^{n\times p}$ with $n$ students as rows and $p$ questions as columns, where each score $X_{is} \in \mathbb{R}_{\geq0}$ is a non-negative real number. For each student $i=1,\ldots,n$, we further denote $X_{i\cdot}=\left(X_{i1},\ldots,X_{i p}\right)^\top\in\mathbb{R}_{\geq0}^p$ as a random vector of student $i$'s scores for the $p$ questions, and $x_{i.}=\left(x_{i1},\ldots,x_{ip}\right)^\mathrm{T}$ is a realization of $X_{i\cdot}$, i.e., student $i$'s actual question scores in a dataset. 

Note that $X_{i\cdot}=\left(X_{i1},\ldots,X_{ip}\right)^\top\in\mathbb{N}^p$ is the random score vector corresponding to student $i,~i=1,\ldots n$. We assume that the random vector $X_{i\cdot}$ jointly follows a $p$-dimensional distribution $F$, independently for $i=1,\ldots,n$. Marginally for each question $s$, we assume that random variables $X_{is}$ independently follow a multinomial distribution $F_s$ with an event probability vector $\theta_s$ and a trial number $1$, i.e., $X_{is}\sim\mathrm{MultiNomial}\left(\theta_s,1\right)$. The event probability vector $\theta_s$ could be estimated by the proportion vector of question scores’ occurrence.

Given the estimated the marginal distributions $\widehat{F_1},\ldots,\widehat{F_p}$, we use a copula to model the joint distribution $F$ of $X_{i\cdot}$. A copula is defined as a cumulative distribution function (CDF) of $p$ random variables, for which marginally each random variable is following a uniform distribution $\mathrm{Uniform}\left[0,1\right]$. Sklar’s theorem states that for continuous marginal CDF distributions $F_s,s=1,...,p$, there exits a unique copula function $C:\left[0,1\right]^p\rightarrow\left[0,1\right]$ such that
\begin{eqnarray*}
F\left(x_{i1},\ldots,x_{ip}\right)=C\left(F_1\left(x_{i1}\right),\ldots,F_p\left(x_{ip}\right)\right),
\end{eqnarray*}
where $x_{i.}=\left(x_{i1},\ldots,x_{ip}\right)^\mathrm{T}$ is a realization of random vector $X_{i\cdot}=\left(X_{i1},\ldots,X_{ip}\right)^\top$. However, in our analysis, the marginal multinomial distributions $F_s$ are not continuous, thus the copula function is not unique, i.e., unidentifiable. We further deploy the distributional transform technique to construct the continuous random variables ${\widetilde{X}}_{is}=X_{is}+V_{is}$, where $V_{is},~i=1,\ldots,n,~s=1,\ldots,p$ are independent random variables following $\mathrm{Uniform}\left[0,1\right]$. We denote the continuous CDF of ${\widetilde{X}}_{is}$ as $\widetilde{F_s}$ and further define
\begin{eqnarray*}
U_{is}=V_{is}F_s\left(X_{is}-1\right)+\left(1-V_{is}\right)F_s\left(X_{is}\right).
\end{eqnarray*}
As shown in our previous paper \cite{sun2021scdesign2}, $U_{is}$ is essentially deploying the CDF $\widetilde{F_s}$ onto the random variable ${\widetilde{X}}_{is}$, i.e., $U_{is}=\widetilde{F_s}\left(X_{is}+V_{is}\right)$. Hence ${\{U}_{is};i=1,\ldots,n\}$ are independent random variables following $\mathrm{Uniform}\left[0,1\right]$. Therefore, the joint CDF $\widetilde{F}$ of the continuous random variables ${\widetilde{X}}_{i\cdot}=\left({\widetilde{X}}_{i1},\ldots,{\widetilde{X}}_{ip}\right)^\top$ has the unique copula expression $\widetilde{F}\left({\widetilde{x}}_{i1},\ldots,{\widetilde{x}}_{ip}\right)=C\left(u_{i1},\ldots,u_{ip}\right)$ according to the Sklar’s theorem, where ${\widetilde{x}}_{is}$ and $u_{is}$ are realizations of ${\widetilde{X}}_{is}$ and $U_{is}$. We thus use $\widetilde{F}$’s unique copula expression $C\left(u_{i1},\ldots,u_{ip}\right)$ to approximate the discrete random variables’ joint CDF $F$:
\begin{eqnarray*}
F\left(x_{i1},\ldots,x_{ip}\right)\approx C\left(u_{i1},\ldots,u_{ip}\right).
\end{eqnarray*}

Specifically, we use a Gaussian copula structure as an approximation of $C\left(u_{i1},\ldots,u_{ip}\right)$ in our analysis, hence the CDF $F$ can be expressed through the following formula:
\begin{eqnarray*}
F\left(x_{i1},\ldots,x_{ip}\right)\approx\Phi_p\left(\Phi^{-1}\left(u_{i1}\right),\ldots,\Phi^{-1}\left(u_{ip}\right);\mathbf{R}\right),
\end{eqnarray*}
where $\Phi_p$ is the CDF of a $p$-dimensional Gaussian distribution with a zero mean vector and a covariance matrix $\mathbf{R}$, and $\Phi$ is the CDF of a standard Gaussian distribution. The copula covariance matrix $\mathbf{R}$ is equal to the correlation matrix of $\left(\Phi^{-1}\left(U_{i1}\right),\ldots,\Phi^{-1}\left(U_{ip}\right)\right)^\top$. To estimate the covariance matrix $\mathbf{R}$, we (1) sample $v_{is}$ from $\mathrm{Uniform}\left[0,1\right]$ independently for $i=1,\ldots,n$ and $s=1, \ldots,p$; (2) calculate quantities $u_{is}$ as 
\begin{eqnarray*}
u_{is}=v_{is}{\widehat{F}}_s\left(x_{is}-1\right)+\left(1-v_{is}\right){\widehat{F}}_s\left(x_{is}\right),
\end{eqnarray*}
where ${\widehat{F}}_s$ is the estimated marginal CDF function for the s-th question; (3) use the sample covariance matrix $\widehat{\mathbf{R}}$ of $\left(\Phi^{-1}\left(u_{i1}\right),\ldots,\Phi^{-1}\left(u_{ip}\right)\right)^\top,i=1,\ldots,n$ as the estimator of $\mathbf{R}$.

In summary, we first estimate $p$ marginal multinomial distributions $F_1,..,F_p$ as $\widehat{F}_1,\ldots,\widehat{F}_p$. Then we calculate the distribution transformed quantities $u_{ij}$ and estimate the Gaussian copula covariance matrix $\widehat{\mathbf{R}}$ to estimate the joint distribution $F$ as:
$\widehat{F}\left(x_{i1},\ldots,x_{ip}\right)=\Phi_p\left(\Phi^{-1}\left(u_{i1}\right),\ldots,\Phi^{-1}\left(u_{ip}\right);\widehat{\mathbf{R}}\right)$,
with $\widehat{\Theta}=\left\{{\widehat{\theta}}_1,\ldots,{\widehat{\theta}}_p,\widehat{\mathbf{R}}\right\}$ as the estimated parameters.

\subsubsection{Generation of a synthetic question score matrix}
For each student cohort $k$, given the fitted $p$-dimensional joint distribution ${\widehat{F}}^{\left(k\right)}$ with parameters ${\widehat{\Theta}}^{\left(k\right)}=\{{\widehat{\theta}}_1^{\left(k\right)},\ldots,{\widehat{\theta}}_p^{\left(k\right)},{\widehat{\mathbf{R}}}^{\left(k\right)}\}$, we generate the synthetic question score matrix for $n^{\left(k\right)}$ students through the following procedures: first, we draw $n^{\left(k\right)}$ vectors, denoted by $z_{i\cdot}^{\left(k\right)\prime}\in\mathbb{R}^p,~i=1,\ldots,n^{\left(k\right)}$, independently from the p-dimensional Gaussian distribution $\Phi_p\left(\cdot;{\widehat{\mathbf{R}}}^{\left(k\right)}\right)$; then, for $s=1,\ldots,p$, convert $z_{is}^{(k)’}$ to $x_{is}^{(k)’}$ by setting $x_{is}^{(k)’}$ as the $\Phi\left(z_{is}^{\left(k\right)\prime}\right)$-th quantile of distribution ${{\widehat{F}}_s}^{\left(k\right)}$, i.e., $\mathrm{MultiNomial}\left({\widehat{\theta}}_s^{\left(k\right)},1\right)$. 

Therefore, matrix $\mathbf{X}^{(k)’}=\left(x_{is}^{(k)’} \right)\in \mathbb{N}^{n^{(k)} \times p}$ is the synthetic question score submatrix corresponding to the student cohort $k$. By horizontally concatenating the $K$ synthetic question score matrices $\mathbf{X}^{(k)’}$, we output $\mathbf{X}^{’}=\left[\mathbf{X}^{(1)’} \cdots \mathbf{X}^{(K)’} \right] \in \mathbb{N}^{n \times p}$ as the final synthetic question score matrix.

\subsection{Synthetic data mimics real data}\label{subsec:supp_synthMimic}
To determine if synthetic control exam data mimics question score data in real exams, we examined their similarity: First, for both synthetic and real question-by-student score matrices, we summarize three statistics (mean, variance, and coefficient of variance) marginally for each question and visualize the distributions of questions’ statistics using violin plots; second, we compare the question-question correlation structures between real and synthetic data using heatmaps of Pearson correlation coefficients; and third, we calculate the median integration local inverse Simpson’s index (miLISI) between synthetic and real students in a 2D embedding. 

The figure compares results for two exams: benchmark exam 1 and proctored control STEM101 F17 Exam 2. The violin plots (left) illustrate that the distributions of questions’ statistics between real and synthetic data mimic each other. The heatmaps (right) show that the synthetic data preserve the question-question correlation structure from the real data. Note that the Pearson Correlations between the classes' scores for each question are generally positive due to students' tendencies to perform either well on most questions or poorly on most questions.

To calculate miLISIs, we first vertically concatenate the synthetic and real question score matrices, then perform a dimension reduction procedure regarding the dimension of questions to find the 2D embedding of synthetic and real students. The dimension reduction procedure involves a principal component analysis (PCA) with $30$ components and a Uniform Manifold Approximation and Projection (UMAP) for 2D embedding. The miLISI ranges from $1$ to $2$, with a value closer to $2$ indicating that the simulated data mix well with the real data in the 2D embeddings. The miLISI for benchmark exam $1$ is $1.938$ and for proctored control STEM101 F17 Exam $2$ is $1.9544$. Both indicate that the data for synthetic students mixes well with that of the real students when embedded into a 2D space.

\begin{figure}[htbp]
\centering
\includegraphics[width=0.9\textwidth]{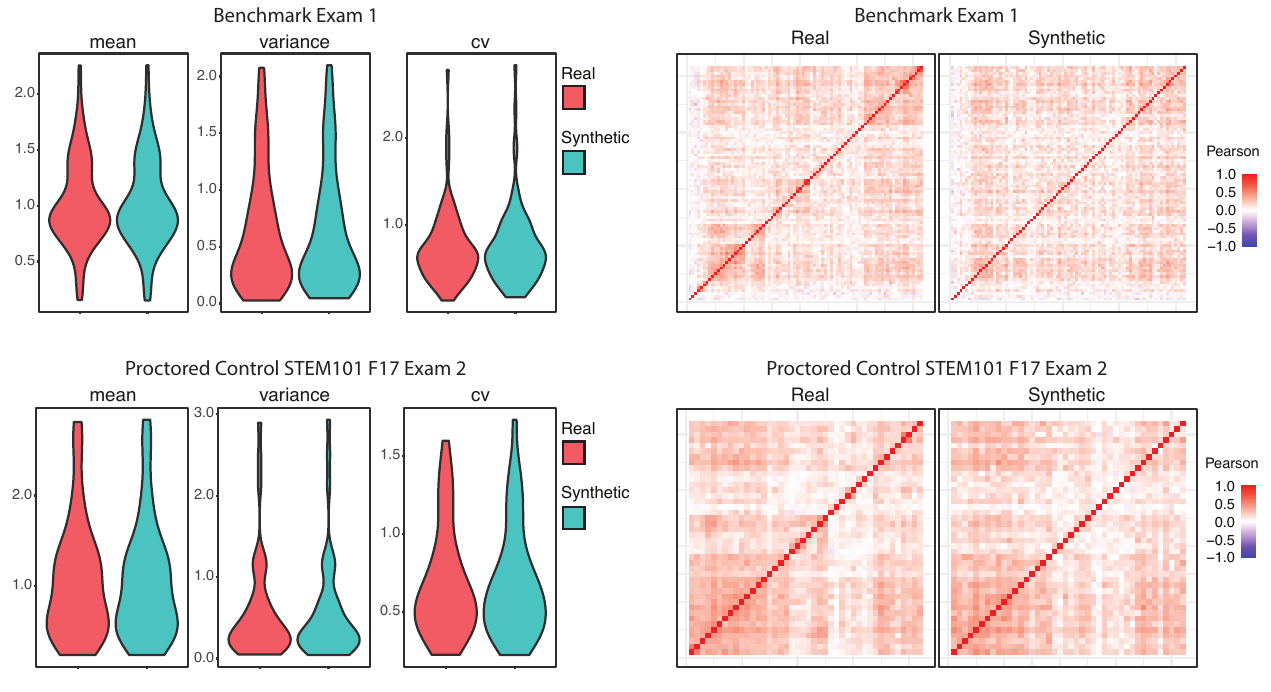}
\caption{Synthetic control exams have similar patterns of question scores as real exams. \label{fig:supp_SynReal}}
\end{figure}

\clearpage
\subsection{Synthetic FPRs are not strongly affected by question number}\label{subsec:supp_synFPR_compare}
The eight proctored control exams that have $>60$ questions were each randomly down-sampled to produce virtual exams with $30$, $40$, $50$ or $60$ questions. Sets of $50$ virtual exams were produced for each of the four numbers of question for each of the proctored controls. Synthetic control exams were generated for each set to give $100{,}000$ student tests, which were then used to determine synthetic FPRs. No systematic change in synthetic FPR is seen with question number (see Fig.~S2). Note that virtual exams containing $20$ questions derived from these eight proctored controls have a complexity $<8$, and thus are not appropriate for Q-SID analysis, see Section \ref{subsec:TPR_comp}.  

\begin{figure}[htbp]
\centering
\includegraphics[width=0.6\textwidth]{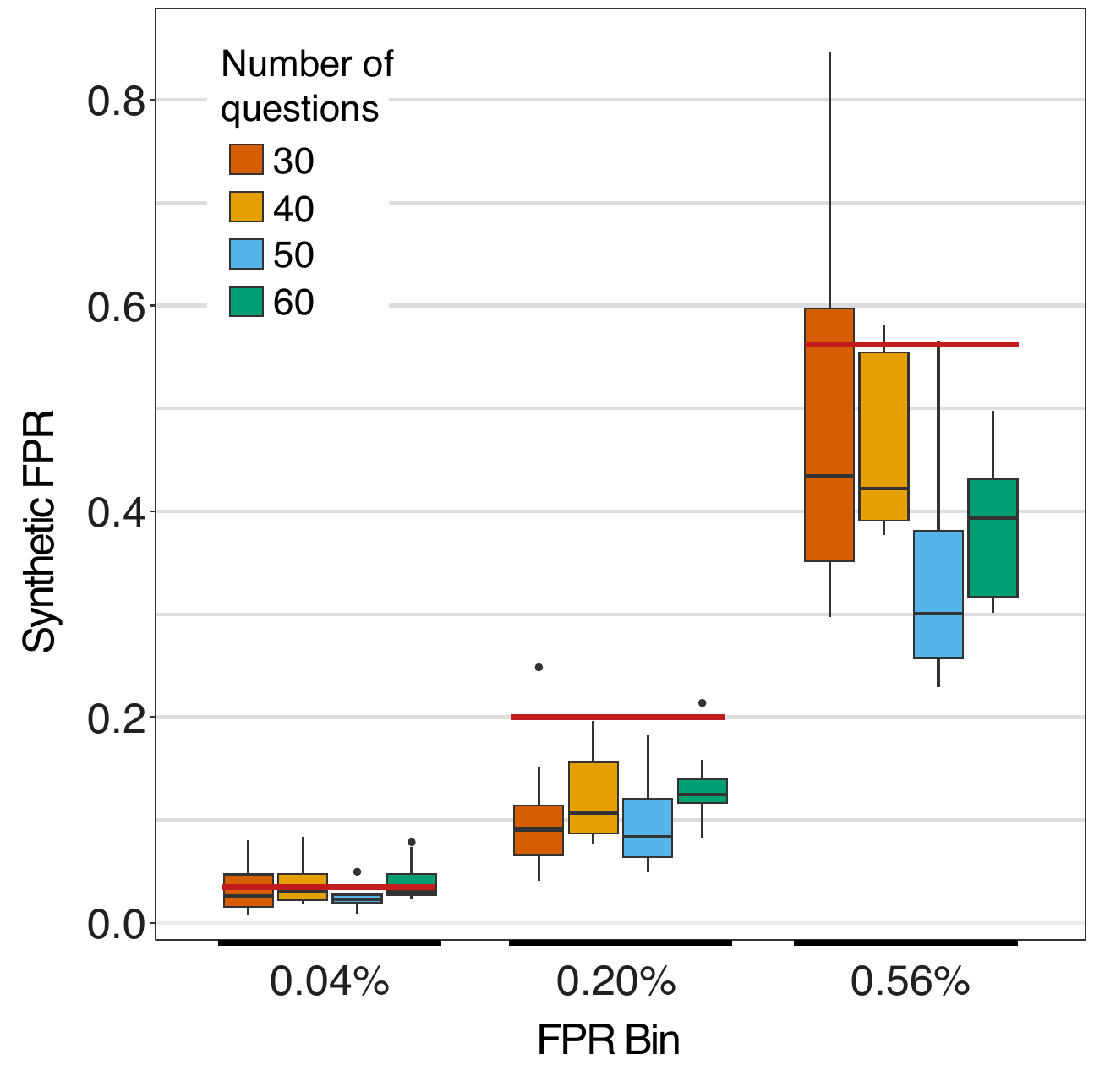}
\caption{Synthetic FPRs show no systematic change with question number (bar and whisker plots). The empirical FPRs are shown by horizontal red lines. \label{fig:supp_synFPR}}
\end{figure}

\clearpage
\section{Example output of Q-SID}\label{sec:supp_output}

\subsection{Example of collusion groups identified in the benchmark exams}\label{subsec:supp_outtable}
The following figure shows the collusion groups identified in the two benchmark exams. Q-SID also outputs the corresponding empirical and synthetic FPRs assigned for each collusion group. Note that the values of the empirical FPRs are constant for all
exams (i.e. $0.04\%$, $0.20\%$ and $0.56\%$), whereas those for the synthetic FPRs vary between
exams. 

\begin{figure}[htbp]
\centering
\includegraphics[width=0.8\textwidth]{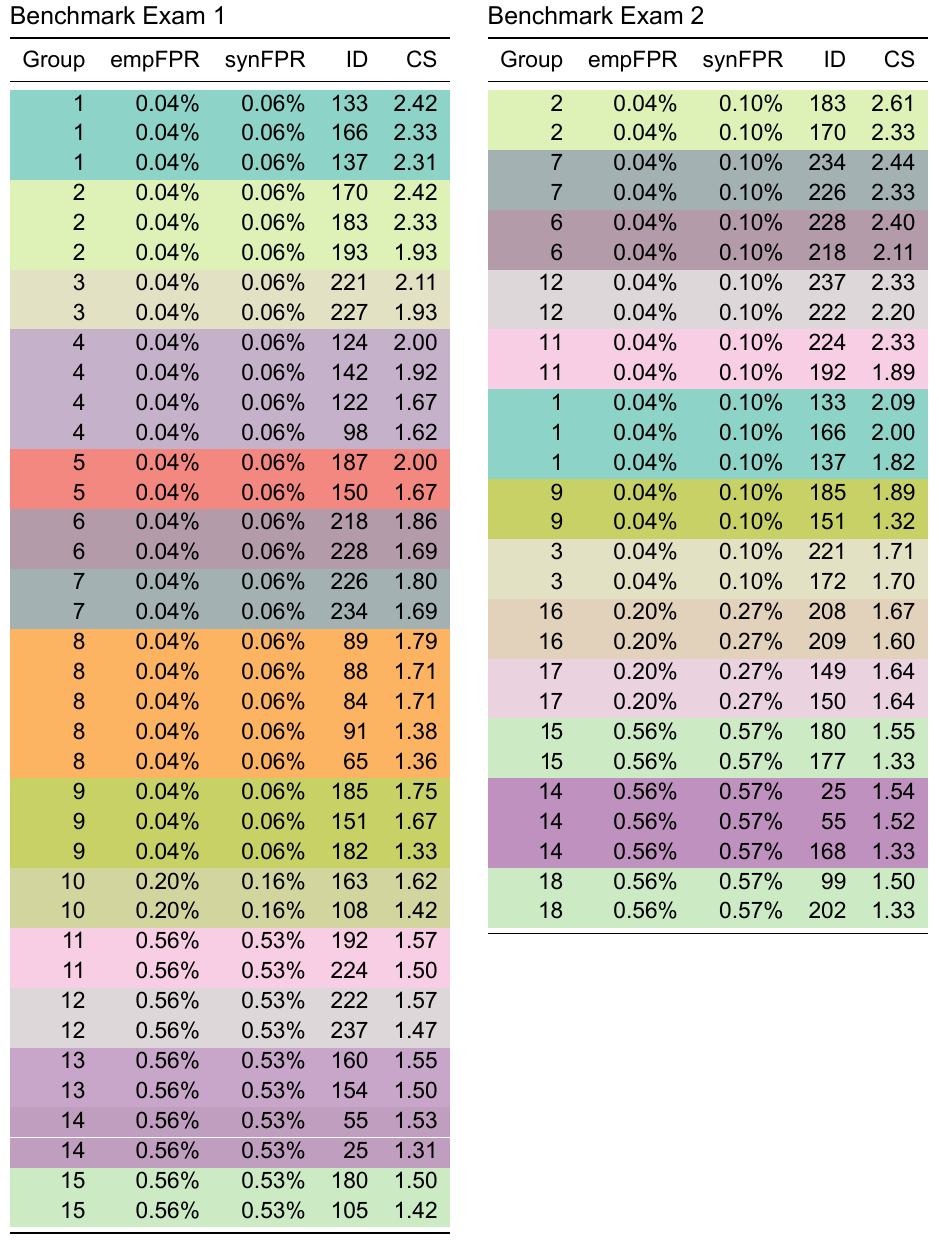}
\caption{Empirical and synthetic FPRs for collusion groups identified in the benchmark exams along with anonymous student IDs and CSs. All students listed are validated true positives who colluded in the groups shown except for students 182 and 168, who did not collude and thus are false positives.  \label{fig:supp_FPRs_table}}
\end{figure}

\clearpage
\subsection{Example Q-SID report}\label{subsec:supp_report}
The Q-SID website produces a report pdf file summarizing the results of the analysis of uploaded query exam data. An example report for benchmark exam $1$ is provided on the following pages. 




\newpage
\bibliographystylelatex{agsm}
\bibliographylatex{gR2_main}

\end{document}